\documentclass{article}
\usepackage{blindtext}
\usepackage[a4paper, top=2cm, bottom=3cm, left=1.5cm, right=1.5cm]{geometry}
\usepackage[english]{babel}
\usepackage{algorithmic}
\usepackage{textcomp}
\usepackage[algoruled,vlined,linesnumbered]{algorithm2e} %
\usepackage{amsmath,amssymb,amsfonts,bm,subcaption}
\usepackage{makecell}
\usepackage{booktabs}
\usepackage{multirow}
\usepackage{float}
\usepackage{graphicx}
\usepackage{indentfirst}
 \usepackage[backend=biber,style=nature]{biblatex}
\usepackage[colorlinks=true, allcolors=blue]{hyperref}
\usepackage{color}

\SetKwInput{KwInput}{Input}                
\SetKwInput{KwOutput}{Output}              

\providecommand{\keywords}[1]
{
  \small	
  \textbf{\textit{Keywords---}} #1
}

\title{Reconstructing 3D Neural Hemodynamics using Sparse Ultrasound Localization Microscopy Data}
\author{
Jipeng Yan$^{1,6}$, Oscar Bates$^2$, Jingwen Zhu$^3$, Qingyuan Tan$^3$, Biao Huang$^3$, John Goodwin$^4$, \\ Andriy S. Kozlov$^4$, Chris Dunsby$^5$, Meng-Xing Tang$^{3,6}$
}
\begin{document}
\date{}
\maketitle
\footnotetext[1]{was with Ultrasound Lab for Imaging and Sensing, Department of Bioengineering, Imperial College London, London SW7 2AZ, UK, and is with the State Key Laboratory of Robotics and Systems, Harbin Institute of Technology, Harbin 150001, China.} 
\footnotetext[2]{Department of Earth Science and Engineering, Imperial College London, London SW7 2AZ, UK.}
\footnotetext[3]{Ultrasound Lab for Imaging and Sensing, Department of Bioengineering, Imperial College London, London SW7 2AZ, UK.}
\footnotetext[4]{Laboratory of Auditory Neuroscience and Biophysics, Department of Bioengineering, Imperial College London, London SW7 2AZ, UK.}
\footnotetext[5]{Light Community, Department of Physics, Imperial College London, London SW7 2AZ, UK.}
\footnotetext[6]{Corresponding authors: Meng-Xing Tang (e-mail: mengxing.tang@imperial.ac.uk, Tel.: +44-2075943664), Jipeng Yan (e-mail: jipengyan@hit.edu.cn)}
\begin{abstract}
\normalsize
Ultrasound Localization Microscopy (ULM) has presented great potential in functional imaging, benefiting from its ability to reconstruct deep microvasculature. However, the hemodynamic reconstruction is compromised by sparsity in the ULM data, as a limited number of MB tracks cannot sample the complete speed profile in one vessel.
 Here, we propose to reconstruct hemodynamics using sparse ULM velocity maps by solving a laminar flow model through stochastic variational inference. In addition to vascular geometry and flow velocity maps, the proposed method generates two new ULM maps—a pressure gradient map and a map describing uncertainty of the estimation. By investigating the effect of sparsity in ULM maps on the quantification and visualization of hemodynamics, we demonstrate the effectiveness of the proposed method in dealing with sparse ULM maps via simulations and 3D rat brain imaging. Accurately reconstructing a broad range of hemodynamic parameters and associate uncertanties using sparse ULM data may help detect subtle and dynamic brain activity.

\end{abstract}

\normalsize

\section{Introduction}


Ultrasound Localization Microscopy (ULM)\cite{christensen2014vivo,errico2015ultrafast} is a type of Super-Resolution Ultrasound (SRUS) technique, achieved by localizing and tracking microbubbles (MBs)\cite{christensen2020super,dencks2025review_super} or erythrocytes \cite{naji2024super}. 
The localization approach allows ULM to break the wave diffraction limit and image deep microvasculature by utilizing the high sensitivity of MBs to ultrasound \cite{opacic2018motion,huang2021super}, which has been demonstrated to outperform Computer Tomography Coronary Angiography (CTCA) in the heart \cite{yan2024transthoracic} and shows promise for the assessment of human brain microvasculature pathologies\cite{demene2021transcranial,bureau2025ultrasound}. The tracking can capture hemodynamics in any directions, which is not available by CTCA, Color Doppler, Power Doppler, or SRUS based on shrinking the point spread function (PSF) using temporal coherence \cite{bar2018sushi,yin2024pattern}. Therefore, ULM has been investigated widely for its feasibility in brain vascular imaging \cite{renaudin2022functional,lin2024eibomedicine,kailiangxu2025brain,jones2025Theranostics,lee2026assessing} and clinical applications \cite{xia2024consensus_statement,smith2026quantitative}.


As the number of MBs in the body is many orders of magnitude less than the blood cells, it is not possible for ULM to fully sample the vast vascular space within a realistic acquisition time frame. The limited accumulation of MB signals in a sparse ULM map can lead to biased hemodynamic quantification and visualization  \cite{ hingot2019microvascular, dencks2020assessing,smith2026quantitative}, as the blood flow speed in the cross-section of the blood vessel is generally not constant and the MBs accumulate in a sparse map that does not cover every location in the vessel. Although the increase of MB concentration could alleviate map sparsity, the concentration should not be too high to avoid inference from overlapping MBs and to ensure the accuracy of existing localization techniques \cite{lerendegui2024ultra}, whether model-based \cite{heiles2022performance} or learning-based  \cite{van2020super,liu2020deep,shin2024context}.
 Therefore, ULM data acquisition takes longer than conventional ultrasound imaging to achieve high-saturation vascular maps, and may require repeated acquisitions to reduce inconsistencies in functional ULMs \cite{renaudin2022functional}, which comes at the cost of losing the fidelity to the true temporal dynamics.
Additionally, blood flow at vessel edges is usually slower than that at the center, so MBs are less likely to pass by vessel sides, making reconstructed vessels narrower than the actual vessel. This problem can be more serious when temporal filtering is adopted to remove tissue signals, such as high-pass filtering and Singular Value Decomposition (SVD) \cite{demene2015spatiotemporal}, where slow-moving MBs can also be removed. Therefore, new methods to improve the geometrical and hemodynamic quantification and visualization are needed when ULM maps are sparse.


 Existing techniques deal with sparse ULM maps mainly by using large pixel sizes or smoothing after accumulation to improve the apparent vascular saturation in the maps. Both  ways compromise image resolution, especially when the pixel size or smoothing kernel is larger than the localization accuracy \cite{hingot2021measuring}. The latter can mitigate the above problem, but smoothing is usually done in a manner of convolution with a fixed kernel, such as a Gaussian kernel for ULM count (density) maps or averaging kernel for ULM speed maps. Information that has not been sampled, such as those lost by the temporal filtering, cannot be recovered after smoothing. Some studies \cite{huang2021super,renaudin2022functional,xing2025primatebrain} utilize vesselness filters \cite{frangi1998multiscale}  on ULM count or speed maps to enhance the image. The vesselness filters output a score, derived from eigenvalues of local Hessian analysis, to describe if the local region contains a vessel. Therefore, the vesselness filters are used for segmenting the vessels but cannot be used for reconstructing hemodynamics.


In this study, we demonstrate the feasibility of reconstructing 3D neuronal hemodynamics from sparse ULM speed maps using a flow dynamics model \emph{in silico} and \emph{in vivo}. We integrate the laminar flow model into stochastic variational inference to estimate hemodynamic parameters with the MB trajectories that are grouped from sparse ULM speed maps based on Hessian Analysis. We show the method’s effectiveness by analyzing the impact of map sparsity on hemodynamic quantification and visualization, and demonstrating new ULM maps of flow pressure gradient and hemodynamics estimation uncertainty.

\section{Results}
\begin{figure}
    \centering
   \includegraphics[width=17 cm]{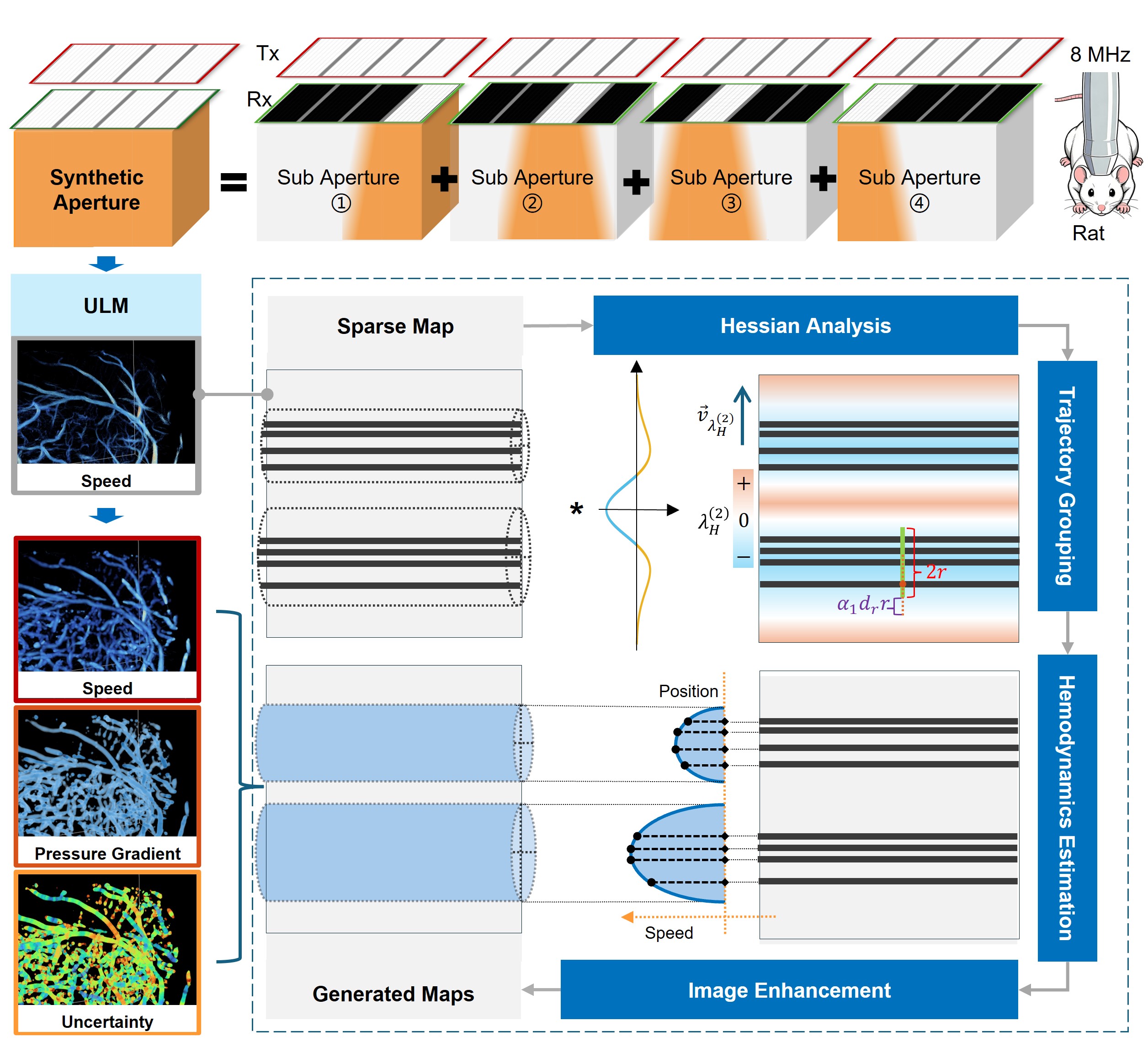}
    \caption{Diagram of using the proposed method, in the dashed box, after a ULM processing pipeline. A ULM speed map is the input of the proposed method. First, the multi-scale Hessian analysis groups MB trajectories on each vessel cross section, then a laminar flow model fits the speeds and coordinates of MB trajectories on a vessel cross section using SVI. This outputs estimated parameters of the laminar flow model as well as the uncertainty of the estimation. By accumulating the estimated speed profile, pressure gradient, or estimation uncertainty on each cross section onto a map the same size as the input ULM speed map, the proposed method can generate three maps: (1) an enhanced speed map, (2) a pressure gradient map, and (3) an uncertainty map. The feasibility is demonstrated by imaging a rat brain with a 1024-elements 8-MHz matrix array probe.}
    \label{fig:1}
\end{figure}

The method proposed to reconstruct hemodynamics from a sparse ULM speed map is shown in Figure \ref{fig:1}. When laminar flow occurs in the majority of vessels, the laminar flow model is adopted to estimate hemodynamic parameters, including radius, center position, pressure gradient and mean velocity, from a group of MB speed values and their positions on a vessel cross-section. The estimation applies Stochastic Variational Inference (SVI) -- the implementation of which can be found in the Methods and Supplementary Method 1 -- to obtain Gaussian distributions of the parameters, where the means are taken as the estimated value and the standard deviations can describe the estimation uncertainties \cite{hoffman2013stochastic,blundell2015weight,bates2022probabilistic}. A strategy based on multi-scale Hessian analysis is adopted to obtain vessel cross-sections and to group speed values and positions on the ULM speed map for the model-based estimation -- the implementation of which can be found in the Methods and Supplementary Method 2. 
Estimated parameters are used to enhance the saturation of the speed map, generate a new map presenting the pressure gradient and create another new map to describe the uncertainty of estimation. The proposed method is evaluated and compared with baseline methods via simulations and demonstrated \emph{in vivo} by imaging a rat brain. 

\subsection{Evaluation with Simulated Data Samples} 

 The proposed model-based method, which solves the Hagen-Poiseuille equation using SVI, is first evaluated and compared with a baseline method, denoted as 'direct', under a simulated situation where trajectories have been exactly grouped for a vessel cross-section. 
 Simulated data, containing various numbers of trajectories, are generated by randomly sampling positions from a round vessel cross-section and giving the speed values with the laminar flow model. The localization uncertainty in ULM processing is simulated by adding Gaussian noise to the coordinates.
  In the direct method, the vessel center and mean speed are estimated using average values of trajectories, the radius is estimated by assuming the trajectory farthest from the center is located at the edge of the vessel.
 In contrast, the outputs of the model-based method are vessel radius, center and pressure gradient. The Hagen–Poiseuille equation is utilized to allow the comparison of parameters estimated by the two methods.

\begin{figure}[!hb]
    \centering
   \includegraphics[width=17 cm]{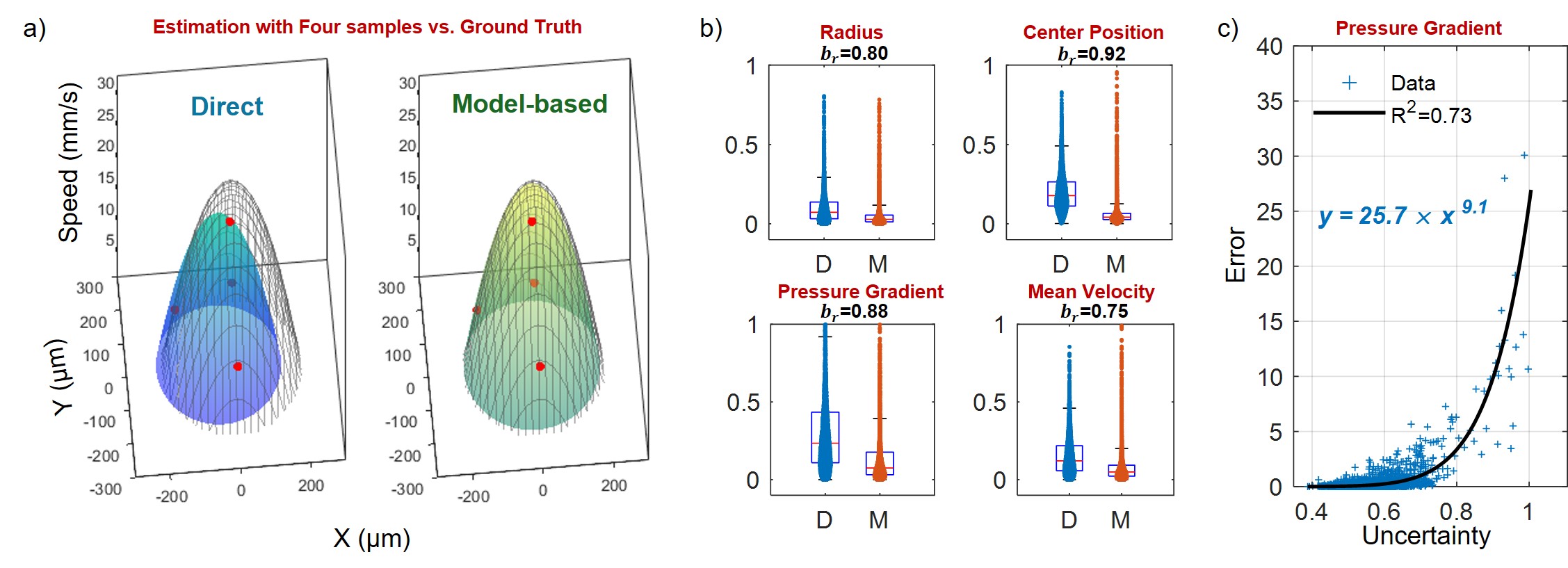}
   \caption{Evaluation on parameter estimation with simulated data samples containing varying numbers of trajectories with different speeds and positions. a) a typical case where four red points are sampled trajectories, colored surfaces are speed profiles created by Hagen–Poiseuille equation with parameters estimated by the two methods, and the gray mesh is ground truth speed profile in the vessel cross section. b) direct estimation errors (denoted as 'D') versus model-based estimation errors (denoted as 'M'), where the proportion (denoted as '$b_r$') by which the latter is lower than the former is given on the top of the corresponding plot. There are 18 (different numbers of trajectories) $\times$ 200 (random sampling repetitions)= 3600 pairs, and Wilcoxon signed rank test implemented by MATLAB \emph{signrank} function returns $p$-value$<$0.001 for each of the four pairs. c) model-based estimation error versus estimation uncertainty with exponential fitting implemented.}
    \label{fig:2}
\end{figure}

A typical case, where four trajectories are at one side of vessels, is shown in Figure \ref{fig:2} a). Here the direct method estimates an off-center flow profile, and the model-based estimation with the flow dynamics model can better handle this situation. The distribution of model-based estimation error versus the direct estimation error is presented in Figure \ref{fig:2} b) with a significant difference found by the Wilcoxon signed rank test, and the proportion of model-based estimation errors below the direct estimation errors over all the estimation is at least 0.75 and can reach 0.92. T
The model-based method can outperform the direct method under a wide range of localization uncertainties, as shown in Supplementary Figure S.1  and S.2 a). Considering that the model-based method might not always perform as expected, the estimation uncertainty obtained by the model-based method is found to be in a positive exponential relationship with the estimation error, as shown in Figure \ref{fig:2} c) and Supplementary Figure S.2 b), indicating the estimation uncertainty can be used to filter out incorrect estimates or illustrate confidence in the estimation. Note that the optimized cost -- Supplementary Method Eq. (S.9) -- in the model-based method cannot be used as such an indicator, because it is difficult to establish a positive correlation between optimization cost and estimation error (Supplementary Figure S.3). The model-based method consistently obtains lower medians of estimation errors than the direct method when the number of trajectories is greater than 3, as shown in Supplementary Figure S.4. The the model-based method requires the determination of four values (radius, two coordinates of the center position, and pressure gradient), so the model-based method is less effective than the direct method when there are only 3 trajectories, as this leads to an ill-posed problem. Therefore, the model-based estimation will not be executed for the following image enhancement when the number of trajectories grouped with Hessian analysis is less than 4.

\subsection{Evaluation with Simulated Images} 
The model-based estimation applied in combination with the trajectory grouping strategy is then evaluated and compared to a smoothing-based method for enhancing the saturation of sparse ULM speed maps of simulated close tubes and a bifurcation vessel. The smoothing-based enhancement is achieved by dividing the average smoothed speed map by the average smoothed density map, which approximates reconstructing the speed map with larger voxels without increasing quantization error.

\begin{figure}[!hb]
    \centering
   \includegraphics[width=13.5 cm]{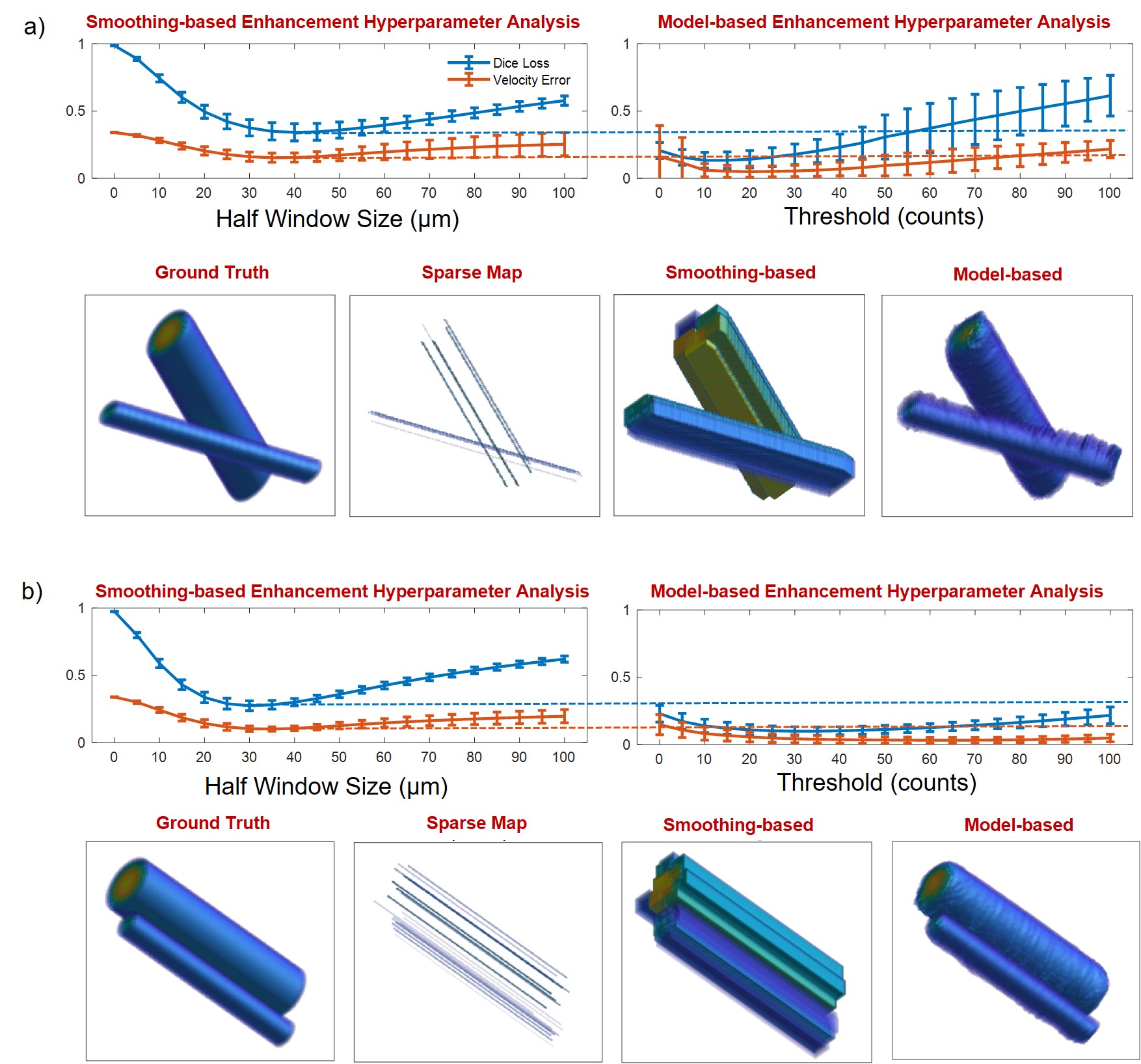}
    \caption{Evaluation on image enhancement with simulated sparse ULM images of two close tubes. a) and b) are obtained with number of trajectories in each tube given as 5 and 10 respectively.
   Curves plot mean and standard deviations of Dice loss and mean square error between the ground truth and speed map enhanced by smoothing using different window sizes for averaging or by model-based estimation using different thresholds to discarding voxels. Images present two cases in which the smoothing-based enhancement achieves the lowest Dice loss with the best window size and the corresponding model-based enhancement images with the best threshold. The 3D volumes are rendered by MATLAB \emph{volshow} function. }
    \label{fig:3}
\end{figure}

\begin{figure}[!hb]
    \centering
   \includegraphics[width=13 cm]{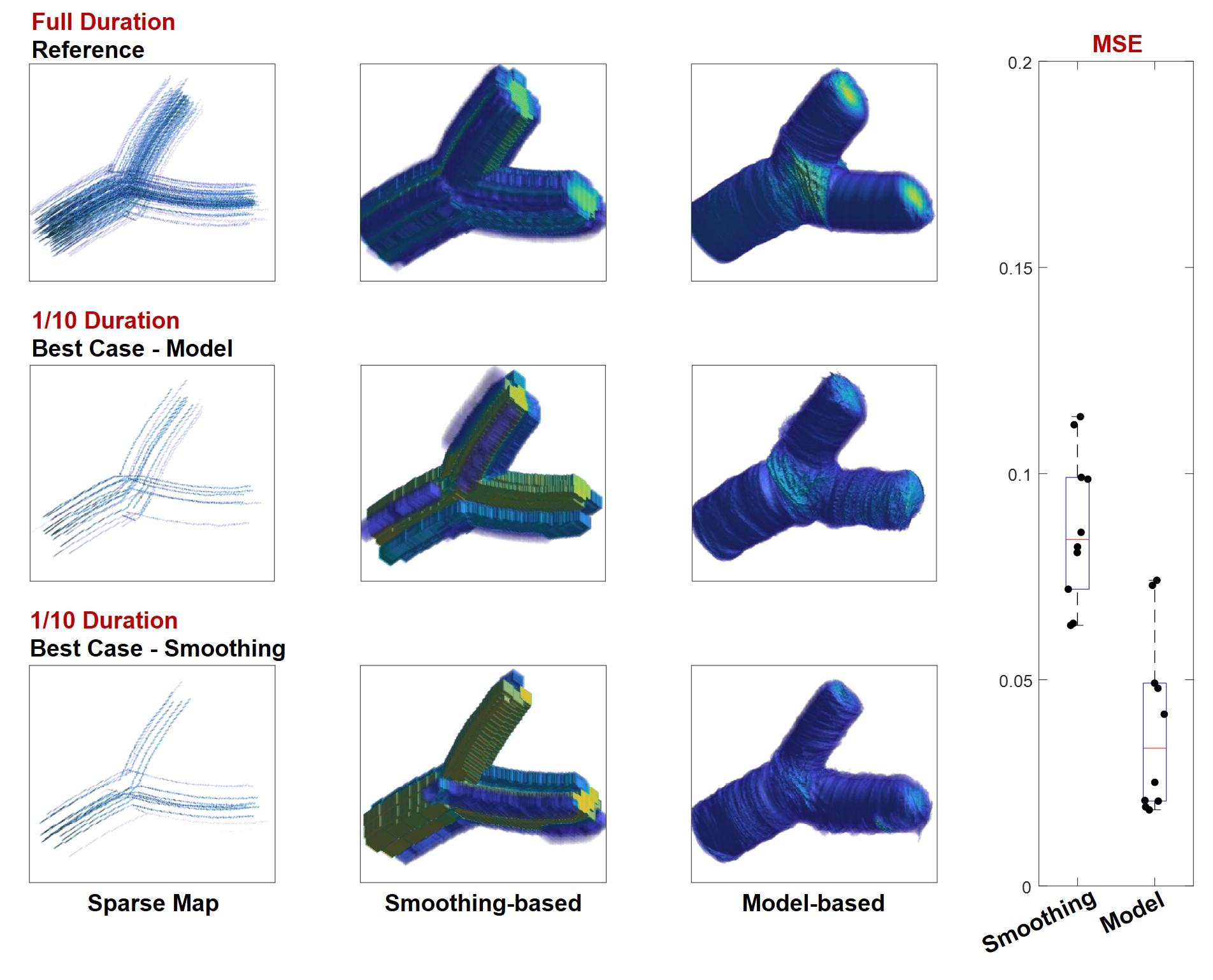}
    \caption{Evaluation on image enhancement with simulated ULM images of a bifurcation vessel. The most suitable hyperparameters obtained in Figure \ref{fig:3} are used here. The images enhanced by two methods with the full-duration data are taken as references respectively.
    Images enhanced by two methods with ten 1/10-duration data are compared with the two references respectively and quantified by normalized mean square error (MSE).  Wilcoxon signed rank test implemented by MATLAB \emph{signrank} function returns $p$-value = 0.002 for two groups of MSE values.
    Two 1/10 data cases are selected as they obtain the least MSE for model-based and smoothing-based enhancement respectively. The 3D volumes are rendered by MATLAB \emph{volshow} function. }
    \label{fig:4}
\end{figure}

Hyperparameters, namely the window size of the average kernel and the count threshold for discarding voxels with low cumulative counts, are applied from small to large find the most suitable parameters for smoothing-based and model-based enhancements respectively. The two methods are compared at their optimal performance, where the enhanced speed map is with least Dice loss to the ground truth. The dashed lines in the Figure \ref{fig:3} a) and b) label the best performance of smoothing-based enhancement averaged across 20 (two simulated angles between tubes and 10 random generations for each angle) ULM images under two different saturation levels. Two enhancement methods obtain lower Dice loss and velocity error with more available trajectories, and the model-based enhancement outperforms the smoothing-based enhancement both in the Dice loss and velocity error. Both cases of enhancement from sparse ULM trajectories demonstrate that model-based enhancements can reconstruct more tubular structures with parabolic velocity distributions in the cross-section. 
Images reconstructed from simulated bifurcation datasets are shown in Figure \ref{fig:4}. The model-based enhancement can achieve smaller mean square error, namely higher similarity, between data acquired with different durations than the smoothing-based enhancement.

\subsection{\emph{In Vivo} Demonstration} 

The feasibility of model-based parameter estimation and image enhancement is validated on ULM imaging of a rat brain, and the differences from direct estimation and smoothing-based enhancement are demonstrated.
The rat brain is imaged by a 1024-elements 8-MHz matrix array that is connected to a 256-channel ultrasound data acquisition system with channel multiplexing.  A plane wave without angle steering is transmitted four times, each of which is followed by receive on sub-apertures with 256 elements. The data are used to synthesize a full aperture when reconstructing ultrasound volumes. B-mode volumes are reconstructed from the received data using a Coherence to Variance (CV) beamformer. Tissue motion due to breathing is detected by rigid image registration, and volumes found with smaller structure similarity index to the preceding volume are  discarded to reduce influence of the pulsatility on the temporal consistence of speed profiles. Tissue signals are reduced by SVD processing in channel signals and then contrast-enhanced ultrasound (CEUS) volumes are reconstructed with the CV beamformer and tissue motion corrected via Lagrangian coordinates. Normalized Cross Correlation (NCC) and feature-motion-model framework are adopted for localizing and tracking MBs, with the tracked trajectories filtered to reduce noise. Straight lines are created between tracked MB positions in consecutive volumes, and speed values are filled to a map with averaging for the voxel that are reached multiple times to create the ULM speed map.

\begin{figure}[!hb]
    \centering
   \includegraphics[width=17 cm]{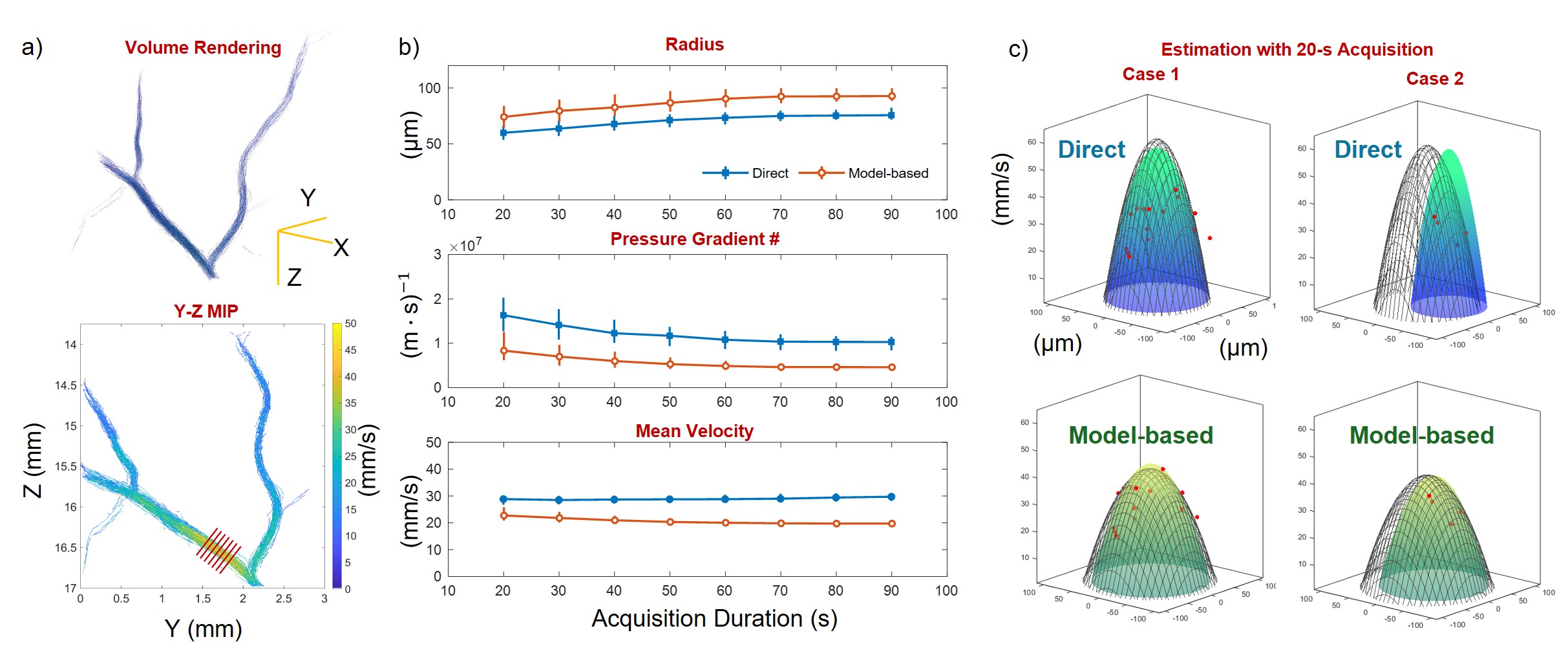}
    \caption{Demonstration on parameter estimation with samples on cross-sections manually picked from a ULM speed map of rat brain. a) Volume rendering by MATLAB \emph{volshow} and Y-Z Maximum Intensity Projection (MIP) of branches under a sub-aperture of the matrix array. 7 cross-sections on a segment presenting straight and relatively unified speed along vessel axial are picked to obtained speeds and coordinates of crossing trajectories. b) Medians, 25th and 75th percentiles of three parameters estimated by the direct and model-based method on samples in speed maps reconstructed from various duration data. Medians and percentiles are calculated across cross-sections and data with the same duration. c) Two cases with most samples and only four samples on a cross section of ULM speed map reconstructed from two different 20-s acquisition data. Red scatters denote the samples' coordinates and speeds, gray meshes denote the speed profile estimated by the two methods with the full 240-s data, and colored surfaces denote the speed profile estimates by the two methods with 20-s data. }
    \label{fig:5}
\end{figure}

Samples on seven manually picked cross-sections are used to demonstrate the difference between the direct and model-based parameter estimation, as shown in Figure \ref{fig:5} a). The model-based method estimates larger radius and lower mean velocity than the direct, as shown in Figure \ref{fig:5} b).  When the model-based method outperforms the direct method in the simulation analysis shown in Figure \ref{fig:2} b), the above results  can be explained by the observation that low-speed MBs traveling near to a vessel edge can be removed with tissue signals by the SVD processing. The direct method can estimate similar parameters and reconstruct similar speed profiles between short and long acquisitions when there are enough data samples, but more differences between estimations of two methods can be observed when too few samples are available, demonstrated by the two cases shown in Figure \ref{fig:5} c). The model-based estimation can  mitigate the problem for the above two cases.

\begin{figure}[p]
    \centering
   \includegraphics[width=17 cm]{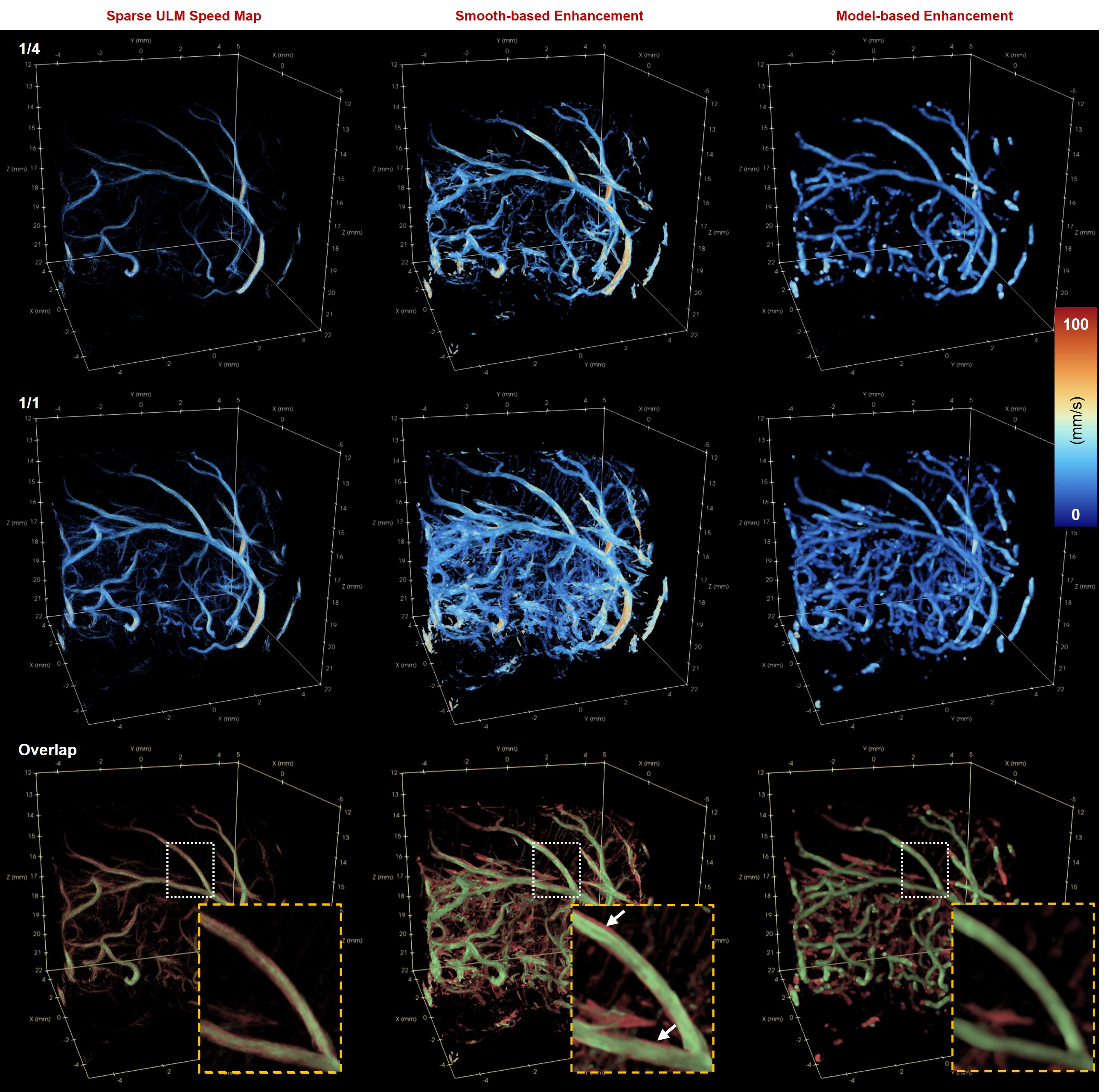}
    \caption{Demonstration on image enhancement with ULM speed map of a rat brain, with 3D maps displayed by ParaView (Version 6.0.0-RC1, Kitware, France). Three columns are volume rendering of ULM speed maps, smoothing-based enhanced speed maps, and model-based enhanced speed map respectively. Three rows are maps obtained with the first quarter of acquisition duration, the full duration, and overlapping the first-row images on the corresponding second-row images with the color changed to green and red respectively. Yellow dashed boxes denote the zoomed-in images of regions in the white dashed boxes. Note that vessels with too few trajectories or too far from the foreground can hardly be visible due to the brightness reduction in the volume rendering processing, but can be visible by Maximum Intensity Projection, as shown in Figure \ref{fig:7}.}
    \label{fig:6}
\end{figure}

The speed maps reconstructed in three ways from short- and long-duration data are shown in Figure \ref{fig:6}. It can be seen that more vasculature can be reconstructed with longer duration data. After overlapping the short-duration map on the corresponding long-duration map, the main difference between the two images reconstructed by the model-based method is the length of reconstructed vessels, because model-based enhancement rejects trajectories grouped for a cross-section when they are too few. The model-based pair presents fewer differences between vessel segments if the segments have already been reconstructed by the short acquisition, such as the arrow-pointed segment, than the other two pairs.

\begin{figure}[p]
    \centering
   \includegraphics[width=17 cm]{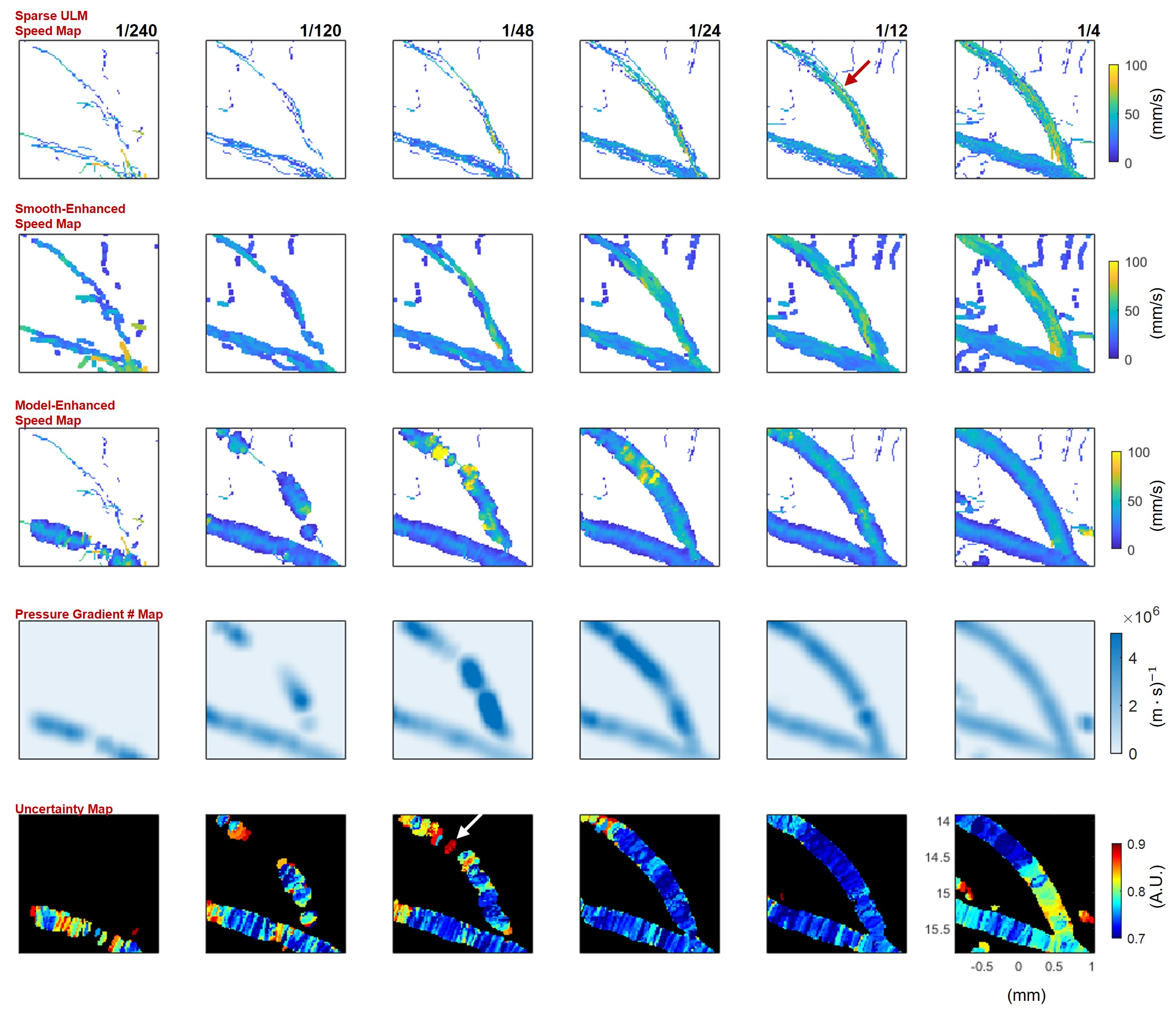}
    \caption{Y-Z MIP of five kinds of maps reconstructed from data with various durations, corresponding to the region of dashed-line box in Figure \ref{fig:6}. First row: Sparse ULM speed maps reconstructed by accumulating trajectories. Second row: Speed maps enhanced by smoothing on the ULM speed maps. Third row: Speed maps enhanced by the model-based method on the ULM speed map. Fourth row: Maps of the division between pressure gradient and blood viscosity estimated by the model-based method from the ULM speed maps.  Fifth row: uncertainty maps generated by the model-based method from the ULM maps for describing the validity of the estimation.}
    \label{fig:7}
\end{figure}

Y-Z MIP images of the white box region in Figure \ref{fig:6} reconstructed with durations from short to long are presented in the columns of Figure \ref{fig:7}. The blank voxels among trajectories in the first row, such as the vessel labeled by the red arrow, can be filled in the images at the second and third rows, demonstrating the benefit of using smoothing or model-based enhancement for improving saturation. There are two main differences between the second and third rows. First, the model-based enhancement presents a more parabolic profile across the vessel than the smoothing-based enhancement. Second, the smoothing-based enhancement dilates all the trajectories, meaning that the saturation enhancement of large vessels is at a cost of overestimating the sizes of small vessels. The model-based enhancement can process trajectories in large and small vessels in different ways. For example, capillaries with small diameters and non-laminar flow can be excluded from model-based estimation, thus maintaining their size unchanged before and after enhancement.

The model-based enhancement can additionally generate a pressure gradient map using a given blood viscosity, which might have the potential for diagnosis. The uncertainty map newly generated by the model-based enhancement can indicate artifacts in the reconstructed image. For example, the high uncertainty region indicated by the white arrow corresponds to a vessel segment with a speed unreasonably higher than that of its neighbors; when more trajectories are available, the speed of the vessel segment becomes similar to that of its neighbors, and the corresponding uncertainty decreases.
\section{Discussion}
This study proposes a model-based method using the laminar flow model for hemodynamic quantification and visualization, demonstrating improved hemodynamic reconstruction from sparse ULM data. The pressure gradient map generated by the proposed method further enriches hemodynamic reconstruction, and the uncertainty map provides new information that indicates the accuracy of the reconstruction.

The method proposed in this study has great potential for brain functional imaging. The quantification with ultrasound Power Doppler is affected by the signal intensity, which varies with the transducer used, the transmit pressure, type of MB and ultrasound attenuation. 
The accuracy of ultrasound Color Doppler relies on a suitable angle between vessel flow and transducer, even when employing the vector Doppler technique \cite{dunmire2000cross}. 
Although binarization in the ULM localization can reduce the influence of signal intensity and the ULM tracking is not sensitive to flow direction, the inconsistency of ULM reconstruction results from the randomness in MB accumulation, which increases with sparsity and can be affected by MB concentrations, acquisition duration, and filtering in the ULM processing. 
The model-based method can achieve higher accuracy in parameter estimation and speed profile reconstruction than the direct estimation, as shown in Figure \ref{fig:2}. The localization uncertainty required to guarantee the outperformance can be achieved \cite{heiles2022performance,shin2024context,dencks2024clinic3D}, and the minimum required sample size is only 4.
When fewer trajectories are required for accurate estimation, the accumulation in ULM can be shortened to capture more dynamic activities.

The proposed model-based image enhancement features with the ability to process ULM images, which are sparser than images of other modalities (such as X-ray, CT, and MRI).
The Hessian analysis that has been widely used for vesselness enhancement is adopted in this study for detecting vessel cross-sections and to group samples from the ULM speed map for model-based estimation. Whereas many existing vesselness enhancement methods require input images with saturated lumen and use eigenvalues to generate a scalar value for each voxel that represents the likelihood of belonging to a tube-like shape \cite{frangi1998multiscale,lamy2021vesselness}, the method presented here only requires input images with sparse trajectories and uses the laminar flow model to generate for each voxel an estimated flow speed and an estimated uncertainty representing the likelihood. Compared to smoothing-based enhancement, the proposed enhancement does not treat trajectories in large and small vessels with a unified dilation, which makes it possible to improve saturation whilst retaining the spatial resolution. 

This study provides a preliminary demonstration of the model-based method -- a method for hemodynamic reconstruction using flow dynamics models -- and has limitations in terms of the number of models studied, the utilization of SVI, and computational efficiency. As only the laminar flow model is used in this study, the enhancement at the location where the blood vessel begins to bifurcate, as shown in Figure 
\ref{fig:4}, is not as accurate as expected.
While SVI is powerful in handling large amounts of data and parameters, to reduce the challenges of convergence, we group a small number of data samples to estimate four values for each cross section independently and primarily use SVI to obtain the uncertainty of the estimate. 
To reduce randomness in image enhancement, the current strategy detects cross sections and a set of trajectories multiple times and calculates the average value at the cost of computational resources.

Future work should be to investigate techniques to further improve the performance of reconstruction and experiments to explore the potential in functional imaging. Modeling the hemodynamic parameters with an assumed distribution that is closer to the real distribution than the Gaussian function could strengthen the correlation between the estimation uncertainty and the estimation error, which will make the estimated uncertainty more effective in indicating estimation errors and filtering estimations. 
Exploring various flow models for different hemodynamics could improve the reconstruction at locations where blood vessels begin to bifurcate, narrow, or widen.  A better strategy, which can accurately group trajectories in a single vessel without the need of repetition, can improve estimation efficiency. It is also worthwhile to explore efficient computational methods, which can fully utilize SVI or other machine learning techniques to reconstruct the entire image simultaneously.
Future research will explore the feasibility of the proposed method in detecting brain activity related to heartbeat \cite{wu20253ddynAmicULM} or stimulation \cite{renaudin2022functional}.



\section{Methods}
\subsection{Implementation of Model-based Parameter Estimation and Image Enhancement}
\subsubsection{Hemodynamic Parameter Estimation with Grouped Trajectories}
The laminar flow on a vessel cross-section can be described by the Hagen–Poiseuille equation below,
\begin{equation}
    v=\frac{G}{4\mu} (R^2-||\boldsymbol{x}-\boldsymbol{x}^c ||^2)
    \label{eq:Poiseuille equation}
\end{equation}
where $G=-dp/dx$ is the pressure gradient;  $\mu$ is the viscosity; $R$ is the radius of vessel; $\boldsymbol{x}^c$ is the vessel center, and $\boldsymbol{x}$ is the location of a passed MB and $v$ is the corresponding speed of passed MB. In this study, we estimate $\hat{G}^\sharp= \frac{\hat{G}}{4\mu}$ for simplicity by avoiding assuming a value for the viscosity.

We estimate posterior distributions of parameters by solving Bayes's equation rather than just estimate the parameters by fitting, because fitting techniques, such as the least square fit between Eq. (\ref{eq:Poiseuille equation}) and the data samples $(\boldsymbol{x},v)$, might output incorrect parameters and provide no indication on its effectiveness.
To obtain a description on the effectiveness, we assume all the data and hemodynamic parameters are in Gaussian distributions and
the Stochastic Variational Inference (SVI) method is utilized to find the posterior distribution of the parameters, given the observed data $\boldsymbol{D}=\{d\}=\{(\boldsymbol{x},v)\}$. 
The vessel mean blood speed can be calculated as $\hat{v}=\hat{G}^\sharp R^2/2$ with the distribution centers, i.e. the mean. Explanation and implementation of SVI for hemodynamic parameter estimation are described in the Supplementary Methods.
Note that Eq. (\ref{eq:Poiseuille equation}) is rewritten in $v=a\boldsymbol{x}^2+b\boldsymbol{x}+c$ for the Adam optimizer used in SVI, as \{$a$, $b$, $c$\} are found easier to converge than \{$G^\sharp$, $R$, $\boldsymbol{x}^c$\}, and \{$G^\sharp$, $R$, $\boldsymbol{x}^c$\} are then calculated with the mean value of \{$a$, $b$, $c$\}, where only $G^\sharp$ keeps unmodified before and after the reparameterization. Therefore,  the standard deviation corresponding to estimated distribution of $a$ is given as the uncertainty in Figure \ref{fig:2} c) to present its correlation to the $G^\sharp$ error, and the geometric mean of standard deviations of $a$, $b$, $c$ is used to present the overall uncertainty of estimation in Figure \ref{fig:7}.



\subsubsection{Trajectory Grouping from ULM Map}
 We assume vessels in ULM images are tubular in shape and group trajectories for the model-based estimation with the algorithm presented in the Supplementary Methods. By convolving the ULM flow map using second-order Gaussian derivatives of different sizes, the cross-section of the blood vessel can be detected using Hessian analysis. Similar with the multi-scale vesselness filter \cite{frangi1998multiscale}, the eigenvectors $\boldsymbol{v}_{\lambda_{H}^{(1)}}$ corresponding to the eigenvalue $\lambda^{(1)}_H$ with least absolute of a local Hessian Matrix regarded along the vessel and the sign of eigenvalues, $\lambda_{H}^{(2)}$ and $\lambda_{H}^{(3)}$, are used to justify if the used Gaussian derivatives is valid to include the voxel in vessel area. In contrast to the vesselness filter, a probability description on a voxel belonging to vessel is not created with the eigenvalues, but trajectories within the vessel cross section, whose direction is defined by the eigenvectors and range is defined by search radius, are grouped as data samples for model-based estimation. Implementation details can be found in the Supplementary Methods.

\subsubsection{Map Generation with Estimated Parameters}
To obtain a saturation-enhanced speed map, a blank map is created in the same size with the original ULM speed map $\boldsymbol{M}$. The speed value for a voxel in the cross section is first calculated with Eq. (\ref{eq:Poiseuille equation}) and parameters estimated by the model-based method for the cross-section, and then accumulated on eight voxels that are in the blank map and surround the voxel in the cross section. The saturation-enhanced speed map is obtained by element-wise division between the accumulated values and the accumulated counts of all the estimated $\boldsymbol{m}$ mean.
 Voxels where the counts are below a threshold $c_T$ are set as zero to reach a good balance between precision and recall in the reconstruction. 
 A pressure gradient map and an estimation uncertainty map are created in a similar way, where the voxel value in the cross section corresponds to the $G^\sharp$ mean and geometric mean of the all the estimated standard deviations. The pressure gradient map is smoothed to reduce discontinuities caused by using the same $G^\sharp$ mean value across the entire cross-section. For the simulation evaluation, enhanced speed maps are created only by the proposed method; for the \emph{in vivo} demonstration, the enhanced speed maps are created by additionally copying the voxels in the original map to corresponding voxels that are with accumulation counts under the threshold $c_T$.




\subsection{Evaluation through Simulations} 
\subsubsection{Parameter Estimation}
To simulate MBs passing through the cross-section of a blood vessel, the positions of the MBs within the disk are randomly sampled using a probability distribution function given by Eq. (\ref{eq:Poiseuille equation}), where MB concentration is assumed constant, and the speed of each MB is also given by the sampled position and Eq. (\ref{eq:Poiseuille equation}). Here, the radius is given as $R_s= 200$ $\mu$m; $\boldsymbol{x}^c_s$ is given as coordinate origin, i.e., (0, 0); maximum speed is given as 30 mm/s and mean speed is $\overline{v}_s= 15$ mm/s, and thus $G^{\sharp}_s=7.5\times 10^{5}$ (m$\cdot$s)$^{-1}$. The number of MB ranges from 3 to 20, with a step size of 1. Gaussian noise with standard deviation of from 0 to $50 \times 10^{-6}$ $\mu$m is added to the sampled positions to simulate different levels of localization uncertainty after ULM processing. Each setting is randomly sampled 200 times. 
Vessel center $\hat{\boldsymbol{x}}^c$,  radius $\hat{R}$, pressure gradient $\hat{G}^\sharp$ and mean velocity $\hat{v}$ are calculated by the
model-based methods, and the estimation errors are calculated and normalized as below.
\begin{equation}
\begin{aligned}
R_e &= |\hat{{R}}-R_s|/R_s\\
\hat{\boldsymbol{x}}^c_e &=\|\hat{\boldsymbol{x}}^c-\boldsymbol{0}\|/R_s \\
G^\sharp_e &=|\hat{G}^\sharp-G^\sharp_s|/G^\sharp_s \\
\overline{v}_e &= |\hat{{v}}-v_s|/v_s\\
\end{aligned}
  \label{eq:error_calculation}
\end{equation}

Vessel center $\hat{\boldsymbol{x}}^c$,  radius $\hat{R}$, pressure gradient $\hat{G}^\sharp$ and mean velocity $\hat{v}$ are also calculated in the way listed below to provide a baseline for comparison.
\begin{itemize}
    \item Vessel center can be obtained by the average of coordinates of each MB: $\hat{\boldsymbol{x}}^c=\overline{\boldsymbol{x}}$.
    \item Vessel radius can be obtained by the maximum of distance of MBs to the estimated vessel center $\hat{R}=max(||\boldsymbol{x}-\hat{\boldsymbol{x}}^c||_2)$.
    \item Vessel mean blood speed can be obtained by the average of MB speed $\hat{v}= \overline{v}$.
    \item Pressure gradient$^\sharp$ can be calculated by
    $\hat{G}^\sharp= \frac{\hat{G}}{4\mu} = 2\overline{v}/R^2$.
\end{itemize}

\subsubsection{Parallel Tubes}
The two straight tubes are placed parallel to each other or at a 45-degree angle. 
Radii of two vessels are 50 and 100 $\mu$m respectively and the distance between the nearest edges is 25 $\mu$m; the maximum speed of the small and large vessels are 15 mm/s and 30 mm/s; 5 or 10 random tracks are generated with Eq. (\ref{eq:Poiseuille equation}) in each vessels to generate different saturation levels. The pixel size is given as a 5 $\times$ 5 $\times$ 5 $\mu$m$^3$. Each set of parameters was used to generate a simulated vessel 10 times, generating 2 $\times$ 2 $\times$ 10 = 40 simulated data.

Dice loss $d_{loss}$  and normalized mean square error $v_{e(M)}$, defined as below, are used to quantify the performance of the proposed method in enhancing saturation of speed maps.
\begin{equation}
\begin{aligned}
&d_{loss} = 1- 2 \times \frac{ \sum{B_E \cap B_T}}{\sum{B_E} + \sum{B_T}} \\
\end{aligned}
  \label{eq:Map_velocity error}
\end{equation}
where $B_E$ and $B_T$ are the binary maps that are estimated by the proposed method and set as vessel regions, and 
\begin{equation}
\begin{aligned}
&v_{e(M)}= \frac{1}{N_\Omega} \sum_{i\in \Omega}\left(\frac{V_E(i)-V_T(i)}{V_c(i)}\right)^2 \\
&\Omega= B_E \cup B_T
\end{aligned}
  \label{eq:Map_velocity error}
\end{equation}
where $V_E$ and $V_T$ are the estimated and set speed maps; $V_c$ is the speed value on the vessel center line that is closest to voxel $i$; ${N_\Omega}$ is number of voxels in $B_E \cup B_T$. 


The proposed image reconstruction method is also compared to conventional smoothing methods at the aspect of saturation enhancement, with $d_{1oss}$ score and $v_{e(M)}$ as metrics. Average smoothing with a cubic length of 2$h_a$+1 is used to mimic the reconstruction using larger pixel size; 
To avoid the speed decrease due to the smoothing, the average-smoothed speed is voxel-by-voxel normalized with average-smoothed density map.
As for the hyper-parameters, the count threshold $c_T$ for the proposed method is adjusted from 0 to 100 with a step of 5; $h_a$ for the average smoothing is adjusted from 0 to 20 voxel by a step of 1, corresponding to m 0 to 100 $\mu$m with a step of 5 $\mu$m. The methods were compared when their corresponding hyper-parameters used achieve the lowest average $d_{1oss}$ score on the simulated datasets.

\subsubsection{Bifurcation Vessel}
MB trajectories in a bifurcation vessel are generated in the BUFF simulator \cite{lerendegui2022BUff}. The simulated space is a cube with a side length of 1 mm; the radius of the main vessel is 150 $\mu$m, and those of the bifurcated vessels are  119 $\mu$m. The blood viscosity is given as 4 mPa/s, generating speeds at the vessel center as 27.9, 23.6 and 20.8 mm/s. 
Trajectories are generated by the \emph{continuous\_injection} function with the average rate as 200, duration of 2 s, simulated time step as 2 ms, and start time as 2 s. 
The obtained trajectories are then accumulated in a pixel map, whose pixel size is 5 $\times$ 5 $\times$ 5 $\mu$m$^3$, with straight lines filled on voxels between MB positions. The ULM speed map is generated via dividing the accumulated speeds by the accumulated event counts. Data with different time durations (one whole 2-second dataset and ten 0.2-second datasets) are used to reconstruct the ULM images at two different saturation levels.

The model-based method and average smoothing are implemented with the corresponding hyper-parameters obtained in the straight tube simulation. The normalized mean square error $v_{e(M)}$ is used to measure the similarity between speed maps reconstructed from the 2-second dataset and each 0.2-second dataset.

\subsection{Demonstration with Rat Brain Data}
\subsubsection{Animal}
A Sprague Dawley rat (male, 23 weeks, 748 g) was used for the imaging procedure. The experiment complied with the Animals (Scientific Procedures) Act 1986 and was approved by the Animal Welfare and Ethical Review Body of Imperial College London. The animal was induced in an induction chamber in 5 \% isoflurane, and then transferred to a small animal physiological monitoring platform (Harvard Apparatus, Massachusetts, USA). Isoflurane was titrated so the breathing rate reached a target of 80 bpm. Urethane was then injected intraperitoneally (1.35 g/kg body weight, 0.5 g/ml, Sigma-Aldrich) and isoflurane was removed, with subsequent 10 \% top-up doses administered as required to eliminate eye-blink and pedal-withdrawal reflexes. Atropine (0.66 ml/kg, 1 \% w/v) was administered subcutaneously to reduce mucous secretions, and bupivicaine (1.5 mg/kg) was injected under the scalp to provide local analgesia. Rectal temperature was monitored throughout the duration of the experiment and maintained at approximately 37 $^\circ$C. The head was fixed in a stereotaxic frame for the duration of the surgery and recordings. A midline incision was made and soft tissue was removed with a scalpel to expose the skull. The right temporalis muscle was resected and a rectangular unilateral craniotomy was performed using a dental drill, using regular applications of sterile saline to cool and clean the skull. The craniotomy was performed just lateral to the midline, extending immediately posterior to the lambdoid suture, anterior to the coronal suture, and down to the zygomatic arch. Homemade microbubbles were infused through the tail vein (0.1 uL/min/g) during the acquisition. After the experiment, animals were humanely killed by overdose of an anaesthetic drug in accordance with Schedule 1 to the Animals (Scientific Procedures) Act 1986.

\subsubsection{Data Acquisition}
The rat brain is imaged by a Vantage 256 ultrasound research system (Verasonics Inc., Kirkland, WA, USA) and a matrix array probe (Vermon, Tours, France) with 32 $\times$ 32 elements, 7.8 MHz center frequency and a 9.3 mm $\times$ 10.2 mm aperture. A multiplexing technique is used to connect the 1024 elements in the probe to the 256 channels in the research system, where 1024 elements are excited at the same time to transmit plane waves and four sub-apertures, i.e., 256 elements in each, receive signals in sequence for one volume. Image volumes are acquired at a frame rate of 400 Hz without beam steering. Data are acquired and saved simultaneously in blocks, 1 s (400 volumes), for the convenience of data loading and off-line processing.

\subsubsection{ULM Processing}
Tissue signals are removed by applying singular value decomposition (SVD) to channel signals for each 400 frames. B-mode and contrast enhanced ultrasound (CEUS) images are reconstructed by the previously described Coherence to Variances (CV) beamformer \cite{yan2023fast}. Slight head motions in all the volumes are detected by rigid image registration to one B-mode volume in the first saved acquisition and corrected for the corresponding CEUS volumes. When pulsatility affects the vessel size and blood flow speed, the structure similarity index is calculated using MATLAB \emph{'ssim'} function between consecutive B-mode volumes, and volumes with low index are regarded as changing fast due to the pulsatility and not used for ULM reconstruction. Considering the compromise between computation cost and effect of voxel size on the localization uncertainty  \cite{song2018effects}, the demonstration is performed for two volumes with different sizes. 

(1) A small volume around 3 $\times$ 3 $\times$ 8 mm$^3$ is reconstructed for B-mode and CEUS images in a map with a voxel size of 25 $\times$ 25 $\times$ 25 $\mu$m$^3$, which is around 1/8 wavelength, and for SR image in a map with a voxel size of 5 $\times$ 5 $\times$ 5 $\mu$m$^3$; 

(2) A large volume around 10 $\times$ 10 $\times$ 10 mm$^3$ with a voxel size of 100 $\times$ 100 $\times$ 100 $\mu$m$^3$ and 20 $\times$ 20 $\times$ 20 $\mu$m$^3$ for original and SR maps respectively. 

MBs are isolated by 3D normalized cross-correlation between each CEUS volume and the Point Spread Function (PSF) estimated via fitting Gaussian functions to a image that is obtained by averaging 10 manually segmented MB main lobes. Noise is reduced by two empirical thresholds in image intensity and cross-correlation coefficients. MBs are localized by the intensity-weighted centroid of CEUS images with the mask created by the thresholding and tracked by the feature-motion-model framework with a maximum accepted speed of 100 mm/s \cite{yan2022super,yan2024transthoracic}. MBs trajectories were created by linking straight lines between the MB positions that are estimated by updating the localized MB position with motion-model-predicted position, according to the Kalman filtering. MBs with track lengths of less than 5 frames are discarded to further improve the precision of localization and tracking. The remaining MBs are used to create ULM speed maps. ULM images are reconstructed with MB trajectories acquired over various durations. 


\subsubsection{Parameter Estimation}
Consecutive subsets of data with various durations, increasing from 20 to 90 seconds, are sampled from the 240-second data.
 150 subsets are created for each duration by changing the start time of subsets from the first second by a step of 1 second. Note that the number of trajectories at one cross-section can change in data with the duration when the MB concentration in blood flow is not constant across time. Seven cross sections are picked manually from a vessel segment that is located approximately under the center of the corresponding sub-aperture. 
 Radius $R$, the division of pressure gradient over blood viscosity $G^\sharp$, and mean velocity $v_e$ are estimated by the direct and model-based methods with the samples. 

\subsubsection{Image Enhancement}
Acquired data with durations of 1, 2, 5, 10, 20, 60, and 240 seconds and starting from the first second are used to reconstruct ULM speed maps for the large volume. The ULM speed maps are enhanced by the averaged-smoothing with a least window size of 3 voxels (70 $\mu m$) and the model-based method with the least count threshold as 0, namely keeping all the accumulated voxels. 



\small
\vspace{3mm}
\noindent\textbf{Acknowledgements}

J.Y. discloses support for the research of this work by the National Natural Science Foundation of China (grant number 62501193), the National Institute for Health Research i4i (grant number NIHR200972), the Engineering and Physical Sciences Research Council (grant number EP/X033651/1), Self-Planned Task (grant number SKLRS202610C) of State Key Laboratory of Robotics and Systems (HIT). O.B. discloses support by the Engineering and Physical Sciences Research Council (grant number EP/X033651/1). J.Z. discloses support for studentship from the China Scholarship Council. Q.T. discloses support by the Wellcome Bioimaging Technology Development Award (grant number 310835/Z/24/Z). 
B.H. discloses support by the Engineering and Physical Sciences Research Council (grant number UKRI145). A.S.K. discloses support by ARIA through Brain-scale Cell Type-specific Acoustic Neural Interface. M.T. discloses support by the Engineering and Physical Sciences Research Council (grant number EP/X033651/1) and ARIA through Brain-scale Cell Type-specific Acoustic Neural Interface.

\vspace{3mm}
\noindent\textbf{Author Contributions}

J.Y., M.T. and C.D. conceived the study. J.Y. designed the algorithm framework and developed codes, and processed the data. O.B. contributed to math and development of SVI. J.Z., J.G. and A.S.K. acquired the \emph{in vivo} data. Q.T. and B.H. contributed to development of image enhancement. J.Y. and M.T. interpreted the results and wrote the first draft. All authors edited and approved the submitted version of the manuscript.

\vspace{3mm}
\noindent\textbf{Data Availability}

The data generated or analyzed during this study are available from the corresponding author upon reasonable request.

\vspace{3mm}
\noindent\textbf{Code Availability}

Codes for the hemodynamic reconstruction are available from the corresponding author upon reasonable request. MB localization, tracking and accumulation in this study is a 3D version of the SRUS Software provided on GitHub (\url{https://github.com/JipengYan1995/SRUSSoftware}). 3D ULM imaging and hemodynamic reconstruction will be integrated into the SRUS Software in the future.

\vspace{3mm}
\noindent\textbf{Competing Interests}

M.T. is a shareholder of Sonalis Imaging Ltd and CardioACC Ltd, and serves as a member of the Scientific Advisory Board for Verasonics inc. O.B. declare an interest in Sonalis Imaging Ltd (stocks, shares and employment). Sonalis Imaging Ltd develops brain ultrasound computed tomography systems; CardioACC Ltd develops intracardiac ultrasound imaging systems; and Verasonics Inc. produces research ultrasound systems.
The other authors declare no competing interests.

 \printbibliography

@article{song2018effects,
  title={On the effects of spatial sampling quantization in super-resolution ultrasound microvessel imaging},
  author={Song, Pengfei and Manduca, Armando and Trzasko, Joshua D and Daigle, Ronald E and Chen, Shigao},
  journal={IEEE Transactions on Ultrasonics, Ferroelectrics, and Frequency Control},
  volume={65},
  number={12},
  pages={2264--2276},
  year={2018},
  publisher={IEEE}
}

@article{smith2026quantitative,
  title={Quantitative image markers of super-resolution ultrasound},
  author={Smith, Cameron AB and Wilson, Harvey and Yan, Jipeng and Tang, Meng-Xing},
  journal={EBioMedicine},
  volume={124},
  year={2026},
  publisher={Elsevier}
}

@article{lerendegui2022BUff,
  title={BUbble flow field: A simulation framework for evaluating ultrasound localization microscopy algorithms},
  author={Lerendegui, Marcelo and Riemer, Kai and Wang, Bingxue and Dunsby, Christopher and Tang, Meng-Xing},
  journal={arXiv preprint arXiv:2211.00754},
  year={2022}
}

@article{heiles2022performance,
  title={Performance benchmarking of microbubble-localization algorithms for ultrasound localization microscopy},
  author={Heiles, Baptiste and Chavignon, Arthur and Hingot, Vincent and Lopez, Pauline and Teston, Eliott and Couture, Olivier},
  journal={Nature Biomedical Engineering},
  volume={6},
  number={5},
  pages={605--616},
  year={2022},
  publisher={Nature Publishing Group UK London}
}

@article{shin2024context,
  title={Context-aware deep learning enables high-efficacy localization of high concentration microbubbles for super-resolution ultrasound localization microscopy},
  author={Shin, YiRang and Lowerison, Matthew R and Wang, Yike and Chen, Xi and You, Qi and Dong, Zhijie and Anastasio, Mark A and Song, Pengfei},
  journal={Nature Communications},
  volume={15},
  number={1},
  pages={2932},
  year={2024},
  publisher={Nature Publishing Group UK London}
}

@article{hingot2019microvascular,
  title={Microvascular flow dictates the compromise between spatial resolution and acquisition time in Ultrasound Localization Microscopy},
  author={Hingot, Vincent and others},
  journal={Scientific Reports},
  volume={9},
  number={1},
  pages={1--10},
  year={2019},
  publisher={Nature Publishing Group}
}

@article{wu20253ddynAmicULM,
  title={3D transcranial Dynamic Ultrasound Localization Microscopy in the mouse brain using a Row-Column Array},
  author={Wu, Alice and Por{\'e}e, Jonathan and Ramos-Palacios, Gerardo and Ghigo, Nin and Bourquin, Chlo{\'e} and Leconte, Alexis and Xing, Paul and Sadikot, Abbas F and Chass{\'e}, Micha{\"e}l and Provost, Jean},
  journal={IEEE Transactions on Biomedical Engineering},
  year={2025},
  publisher={IEEE},
    volume={73},
  number={3},
  pages={1112-1123},
}

@article{kailiangxu2025brain,
  title={Brain-Wide Transcranial Ultrasound Localization Microscopy of the Non-Human Primate},
  author={Guo, Yuanyang and Sun, Qiandong and Xie, Yang and Minonzio, Jean-Gabriel and Xu, Kailiang and Ta, Dean},
  journal={IEEE Transactions on Ultrasonics, Ferroelectrics, and Frequency Control},
  year={2025},
  publisher={IEEE},
    volume={72},
  number={11},
  pages={1448-1461},
}

@article{naji2024super,
  title={Super-resolution ultrasound imaging using the erythrocytes—Part II: Velocity images},
  author={Naji, Mostafa Amin and Taghavi, Iman and Schou, Mikkel and Pr{\ae}sius, Sebastian Kazmarek and Hansen, Lauge Naur and Panduro, Nathalie Sarup and Andersen, Sofie Bech and S{\o}gaard, Stinne Byrholdt and Gundlach, Carsten and Kjer, Hans Martin and others},
  journal={IEEE Transactions on Ultrasonics, Ferroelectrics, and Frequency Control},
  volume={71},
  number={8},
  pages={945--959},
  year={2024},
  publisher={IEEE}
}

@article{hoffman2013stochastic,
  title={Stochastic variational inference},
  author={Hoffman, Matthew D and Blei, David M and Wang, Chong and Paisley, John},
  journal={Journal of Machine Learning Research},
  year={2013}
}

@article{lerendegui2024ultra,
  title={ULTRA-SR challenge: Assessment of ultrasound localization and tracking algorithms for super-resolution imaging},
  author={Lerendegui, Marcelo and Riemer, Kai and Papageorgiou, Georgios and Wang, Bingxue and Arthur, Lachlan and Chavignon, Arthur and Zhang, Tao and Couture, Olivier and Huang, Pingtong and Ashikuzzaman, Md and others},
  journal={IEEE transactions on medical imaging},
  volume={43},
  number={8},
  pages={2970--2987},
  year={2024},
  publisher={IEEE}
}

@article{liu2020deep,
  title={Deep learning for ultrasound localization microscopy},
  author={Liu, Xin and Zhou, Tianyang and Lu, Mengyang and Yang, Yi and He, Qiong and Luo, Jianwen},
  journal={IEEE transactions on medical imaging},
  volume={39},
  number={10},
  pages={3064--3078},
  year={2020},
  publisher={IEEE}
}

@article{xia2024consensus_statement,
  title={Super-resolution ultrasound and microvasculomics: a consensus statement},
  author={Xia, ShuJun and Zheng, YuHang and Hua, Qing and Wen, Jing and Luo, XiaoMao and Yan, JiPing and Bai, BaoYan and Dong, YiJie and Zhou, JianQiao},
  journal={European radiology},
  volume={34},
  number={11},
  pages={7503--7513},
  year={2024},
  publisher={Springer}
}

@article{jones2025Theranostics,
  title={Non-invasive volumetric ultrasound localization microscopy detects vascular changes in mice with Alzheimer's disease},
  author={Jones, Rebecca M and DeRuiter, Ryan M and Deshmukh, Mohanish and Dayton, Paul A and Pinton, Gianmarco F},
  journal={Theranostics},
  volume={15},
  number={3},
  pages={1110},
  year={2025}
}

@article{lin2024eibomedicine,
  title={Super-resolution ultrasound imaging reveals temporal cerebrovascular changes with disease progression in female 5$\times$ FAD mouse model of Alzheimer's disease: correlation with pathological impairments},
  author={Lin, Haoming and Wang, Zidan and Liao, Yingtao and Yu, Zhifan and Xu, Huiqin and Qin, Ting and Tang, Jianbo and Yang, Xifei and Chen, Siping and Chen, Xin and others},
  journal={EBioMedicine},
  volume={108},
  year={2024},
  publisher={Elsevier}
}

@article{xing2025primatebrain,
  title={3D ultrasound localization microscopy of the nonhuman primate brain},
  author={Xing, Paul and Perrot, Vincent and Dominguez-Vargas, Adan Ulises and Por{\'e}e, Jonathan and Quessy, Stephan and Dancause, Numa and Provost, Jean},
  journal={EBioMedicine},
  volume={111},
  year={2025},
  publisher={Elsevier}
}

@article{bureau2025ultrasound,
  title={Ultrasound matrix imaging for 3D transcranial in vivo localization microscopy},
  author={Bureau, Flavien and Denis, Louise and Coudert, Antoine and Fink, Mathias and Couture, Olivier and Aubry, Alexandre},
  journal={Science Advances},
  volume={11},
  number={31},
  pages={eadt9778},
  year={2025},
  publisher={American Association for the Advancement of Science}
}

@article{renaudin2022functional,
  title={Functional ultrasound localization microscopy reveals brain-wide neurovascular activity on a microscopic scale},
  author={Renaudin, No{\'e}mi and Demen{\'e}, Charlie and Dizeux, Alexandre and Ialy-Radio, Nathalie and Pezet, Sophie and Tanter, Mickael},
  journal={Nature Methods},
  volume={19},
  number={8},
  pages={1004--1012},
  year={2022},
  publisher={Nature Publishing Group US New York}
}

@inproceedings{frangi1998multiscale,
  title={Multiscale vessel enhancement filtering},
  author={Frangi, Alejandro F and Niessen, Wiro J and Vincken, Koen L and Viergever, Max A},
  booktitle={International conference on medical image computing and computer-assisted intervention},
  pages={130--137},
  year={1998},
  organization={Springer}
}

@article{dencks2024clinic3D,
  title={Ultrasound localization microscopy precision of clinical 3-D ultrasound systems},
  author={Dencks, Stefanie and Lisson, Thomas and Oblisz, Nico and Kiessling, Fabian and Schmitz, Georg},
  journal={IEEE Transactions on Ultrasonics, Ferroelectrics, and Frequency Control},
  volume={71},
  number={12: Breaking the Resolution Barrier in Ultrasound},
  pages={1677--1689},
  year={2024},
  publisher={IEEE}
}

@article{dencks2020assessing,
  title={Assessing vessel reconstruction in ultrasound localization microscopy by maximum likelihood estimation of a zero-inflated poisson model},
  author={Dencks, Stefanie and others},
  journal={IEEE Transactions on Ultrasonics, Ferroelectrics, and Frequency Control},
  volume={67},
  number={8},
  pages={1603--1612},
  year={2020},
  publisher={IEEE}
}

@article{demene2021transcranial,
  title={Transcranial ultrafast ultrasound localization microscopy of brain vasculature in patients},
  author={Demen{\'e}, Charlie and Robin, Justine and Dizeux, Alexandre and Heiles, Baptiste and Pernot, Mathieu and Tanter, Mickael and Perren, Fabienne},
  journal={Nature Biomedical Engineering},
  volume={5},
  number={3},
  pages={219--228},
  year={2021},
  publisher={Nature Publishing Group}
}

@article{huang2021super,
  title={Super-resolution ultrasound localization microscopy based on a high frame-rate clinical ultrasound scanner: an in-human feasibility study},
  author={Huang, Chengwu and Zhang, Wei and Gong, Ping and Lok, U-Wai and Tang, Shanshan and Yin, Tinghui and Zhang, Xirui and Zhu, Lei and Sang, Maodong and Song, Pengfei and others},
  journal={Physics in Medicine and Biology},
  volume={66},
  number={8},
  pages={08NT01},
  year={2021},
  publisher={IOP Publishing}
}

@article{dencks2025review_super,
  title={Super-resolution ultrasound: from data acquisition and motion correction to localization, tracking, and evaluation},
  author={Dencks, Stefanie and Lowerison, Matthew and Hansen-Shearer, Joseph and Shin, YiRang and Schmitz, Georg and Song, Pengfei and Tang, Meng-Xing},
  journal={IEEE Transactions on Ultrasonics, Ferroelectrics, and Frequency Control},
  year={2025},
  publisher={IEEE}
}

@article{bar2018sushi,
  title={SUSHI: Sparsity-based ultrasound super-resolution hemodynamic imaging},
  author={Bar-Zion, Avinoam and Solomon, Oren and Tremblay-Darveau, Charles and Adam, Dan and Eldar, Yonina C},
  journal={IEEE Transactions on Ultrasonics, Ferroelectrics, and Frequency control},
  volume={65},
  number={12},
  pages={2365--2380},
  year={2018},
  publisher={IEEE}
}

@article{van2020super,
  title={Super-resolution ultrasound localization microscopy through deep learning},
  author={Van Sloun, Ruud JG and Solomon, Oren and Bruce, Matthew and Khaing, Zin Z and Wijkstra, Hessel and Eldar, Yonina C and Mischi, Massimo},
  journal={IEEE transactions on medical imaging},
  volume={40},
  number={3},
  pages={829--839},
  year={2020},
  publisher={IEEE}
}

@article{opacic2018motion,
  title={Motion model ultrasound localization microscopy for preclinical and clinical multiparametric tumor characterization},
  author={Opacic, Tatjana and Dencks, Stefanie and Theek, Benjamin and Piepenbrock, Marion and Ackermann, Dimitri and Rix, Anne and Lammers, Twan and Stickeler, Elmar and Delorme, Stefan and Schmitz, Georg and others},
  journal={Nature Communications},
  volume={9},
  number={1},
  pages={1--13},
  year={2018},
  publisher={Nature Publishing Group}
}

@article{christensen2014vivo,
  title={In vivo acoustic super-resolution and super-resolved velocity mapping using microbubbles},
  author={Christensen-Jeffries, Kirsten and Browning, Richard J and Tang, Meng-Xing and Dunsby, Christopher and Eckersley, Robert J},
  journal={IEEE Transactions on Medical Imaging},
  volume={34},
  number={2},
  pages={433--440},
  year={2014},
  publisher={IEEE}
}

@article{errico2015ultrafast,
  title={Ultrafast ultrasound localization microscopy for deep super-resolution vascular imaging},
  author={Errico, Claudia and Pierre, Juliette and Pezet, Sophie and Desailly, Yann and Lenkei, Zsolt and Couture, Olivier and Tanter, Mickael},
  journal={Nature},
  volume={527},
  number={7579},
  pages={499--502},
  year={2015},
  publisher={Nature Publishing Group}
}

@article{yan2024transthoracic,
  title={Transthoracic ultrasound localization microscopy of myocardial vasculature in patients},
  author={Yan, Jipeng and Huang, Biao and Tonko, Johanna and Toulemonde, Matthieu and Hansen-Shearer, Joseph and Tan, Qingyuan and Riemer, Kai and Ntagiantas, Konstantinos and Chowdhury, Rasheda A and Lambiase, Pier D and others},
  journal={Nature Biomedical Engineering},
  volume={8},
  pages={689--700},
  year={2024},
  publisher={Nature Publishing Group UK London}
}

@article{yin2024pattern,
  title={Pattern recognition of microcirculation with super-resolution ultrasound imaging provides markers for early tumor response to anti-angiogenic therapy},
  author={Yin, Jingyi and Dong, Feihong and An, Jian and Guo, Tianyu and Cheng, Heping and Zhang, Jiabin and Zhang, Jue},
  journal={Theranostics},
  volume={14},
  number={3},
  pages={1312},
  year={2024}
}

@article{christensen2020super,
  title={Super-resolution ultrasound imaging},
  author={Christensen-Jeffries, Kirsten and Couture, Olivier and Dayton, Paul A and Eldar, Yonina C and Hynynen, Kullervo and Kiessling, Fabian and O'Reilly, Meaghan and Pinton, Gianmarco F and Schmitz, Georg and Tang, Meng-Xing and others},
  journal={Ultrasound in Medicine and Biology},
  volume={46},
  number={4},
  pages={865--891},
  year={2020},
  publisher={Elsevier}
}

@article{dunmire2000cross,
  title={Cross-beam vector Doppler ultrasound for angle-independent velocity measurements},
  author={Dunmire, B and Beach, KW and Labs, KH and Plett, M and Strandness Jr, DE},
  journal={Ultrasound in Medicine and Biology},
  volume={26},
  number={8},
  pages={1213--1235},
  year={2000},
  publisher={Elsevier}
}

@article{yan2022super,
  title={Super-resolution ultrasound through sparsity-based deconvolution and multi-feature tracking},
  author={Yan, Jipeng and Zhang, Tao and Broughton-Venner, Jacob and Huang, Pintong and Tang, Meng-Xing},
  journal={IEEE Transactions on Medical Imaging},
  volume={41},
  number={8},
  pages={1938--1947},
  year={2022},
  publisher={IEEE}
}

@article{yan2023fast,
  title={Fast 3D super-resolution ultrasound with adaptive weight-based beamforming},
  author={Yan, Jipeng and Wang, Bingxue and Riemer, Kai and Hansen-Shearer, Joseph and Lerendegui, Marcelo and Toulemonde, Matthieu and Rowlands, Christopher J and Weinberg, Peter D and Tang, Meng-Xing},
  journal={IEEE Transactions on Biomedical Engineering},
  volume={70},
  number={9},
  pages={2752--2761},
  year={2023},
  publisher={IEEE}
}

@article{demene2015spatiotemporal,
  title={Spatiotemporal clutter filtering of ultrafast ultrasound data highly increases Doppler and fUltrasound sensitivity},
  author={Demen{\'e}, Charlie and Deffieux, Thomas and Pernot, Mathieu and Osmanski, Bruno-F{\'e}lix and Biran, Val{\'e}rie and Gennisson, Jean-Luc and Sieu, Lim-Anna and Bergel, Antoine and Franqui, Stephanie and Correas, Jean-Michel and others},
  journal={IEEE Transactions on Medical Imaging},
  volume={34},
  number={11},
  pages={2271--2285},
  year={2015},
  publisher={IEEE}
}

@article{lee2026assessing,
  title={Assessing cerebral capillary function and stalling using single capillary reporters in ultrasound localization microscopy},
  author={Lee, Stephen A and Leconte, Alexis and Wu, Alice and Kinugasa, Joshua and Palacios, Gerardo Ramos and Por{\'e}e, Jonathan and Sadikot, Abbas F and Linninger, Andreas and Provost, Jean},
  journal={Proceedings of the National Academy of Sciences},
  volume={123},
  number={2},
  pages={e2509564123},
  year={2026},
  publisher={National Academy of Sciences}
}

@inproceedings{blundell2015weight,
  title={Weight uncertainty in neural network},
  author={Blundell, Charles and Cornebise, Julien and Kavukcuoglu, Koray and Wierstra, Daan},
  booktitle={International conference on machine learning},
  pages={1613--1622},
  year={2015},
  organization={PMLR}
}

@article{bates2022probabilistic,
  title={A probabilistic approach to tomography and adjoint state methods, with an application to full waveform inversion in medical ultrasound},
  author={Bates, Oscar and Guasch, Lluis and Strong, George and Caradoc Robins, Thomas and Calderon-Agudo, Oscar and Cueto, Carlos and Cudeiro, Javier and Tang, Mengxing},
  journal={Inverse Problems},
  volume={38},
  number={4},
  pages={045008},
  year={2022},
  publisher={IOP Publishing}
}

@article{hingot2021measuring,
  title={Measuring image resolution in ultrasound localization microscopy},
  author={Hingot, Vincent and Chavignon, Arthur and Heiles, Baptiste and Couture, Olivier},
  journal={IEEE transactions on medical imaging},
  volume={40},
  number={12},
  pages={3812--3819},
  year={2021},
  publisher={IEEE}
}

@inproceedings{lamy2021vesselness,
  title={Vesselness filters: A survey with benchmarks applied to liver imaging},
  author={Lamy, Jonas and Merveille, Odyss{\'e}e and Kerautret, Bertrand and Passat, Nicolas and Vacavant, Antoine},
  booktitle={2020 25th international conference on pattern recognition (ICPR)},
  pages={3528--3535},
  year={2021},
  organization={IEEE}
}

@article{seeger2004gaussian,
  title={Gaussian processes for machine learning},
  author={Seeger, Matthias},
  journal={International journal of neural systems},
  volume={14},
  number={02},
  pages={69--106},
  year={2004},
  publisher={World Scientific}
}
\end{document}


\title{Supplementary Methods and Figures}
\date{}
\maketitle
\renewcommand{\theequation}{S.\arabic{equation}}
\renewcommand{\thefigure}{S.\arabic{figure}}

\section{Method 1: Parameter Estimation}
Bayes’ equation is only tractable in specific cases, such as the Gaussian Process \cite{seeger2004gaussian}. Stochastic Variational Inference (SVI) allows solutions to Bayes equation where the likelihood distribution is either analytically or computationally intractable. 
The Bayes' equation can be written
\begin{equation}
    p(\boldsymbol{m}|\boldsymbol{D})= \frac{p(\boldsymbol{D}|\boldsymbol{m})p(\boldsymbol{m})}{p(\boldsymbol{D})}
    \label{eq:BayesEquation}
\end{equation}
where $p(\boldsymbol{m}|\boldsymbol{D})$ is the posterior distribution, i.e, parameter distribution given the observed position and locations at the current cross-section; $p(\boldsymbol{m})$ is the prior distribution, i.e., the parameter distribution of all kinds of vessels; $p(\boldsymbol{D})$ is the evidence distribution, i.e. distribution of MBs' position and locations on cross-section of all kinds of vessels; and the conditional distribution $p(\boldsymbol{D}|\boldsymbol{m})$ is the likelihood to get an evidence given one kind of vessel structure, which can be established with the flow dynamic model, such as Eq. (1) in the main text.

When it is challenging to exactly know the above three distributions, the SVI method is used to find an approximation  $q(\boldsymbol{m}|\boldsymbol{D})$ of the posterior $p(\boldsymbol{m}|\boldsymbol{D})$ by minimizing the Kulback–Leibler (KL) divergence,
\begin{equation}
\begin{aligned}
       q(\boldsymbol{m}|\boldsymbol{D}) &= \arg \min_{q(\boldsymbol{m}|\boldsymbol{D})}\mathbf{KL}[q(\boldsymbol{m}|\boldsymbol{D})||p(\boldsymbol{m}|\boldsymbol{D})] \\
    \label{eq:qmD_defination}
\end{aligned}
\end{equation}
where $q(\boldsymbol{m}|\boldsymbol{D})$ is the approximating distribution of the posterior $p(\boldsymbol{m}|\boldsymbol{D})$, we use the symbol $\boldsymbol{\xi}$ to replace $\boldsymbol{m}|\boldsymbol{D}$ for convenience, i.e, $\xi=(G^{\sharp}_{\boldsymbol{\xi}}, R_{\boldsymbol{\xi}}, \boldsymbol{x}^{\sharp}_{\boldsymbol{\xi}})$. The KL divergence equation can be rearranged to as follows,
\begin{equation}
\begin{aligned}
       \mathbf{KL}[q(\boldsymbol{\xi})||p(\boldsymbol{\xi})] &=  \int{q(\boldsymbol{\xi})\log{\frac{q(\boldsymbol{\xi})}{p(\boldsymbol{\xi})}}d\boldsymbol{\xi}}  
      = \int{q(\boldsymbol{\xi})[\log q(\boldsymbol{\xi})-\log{\frac{p(\boldsymbol{D}|\boldsymbol{m})p(\boldsymbol{m})}{p(\boldsymbol{D})}}]d\boldsymbol{\xi}} \\
     & = \int{q(\boldsymbol{\xi})[\log q(\boldsymbol{\xi})-\log p(\boldsymbol{D}|\boldsymbol{m}) - \log p(\boldsymbol{m}) + \log p(\boldsymbol{D})]d\boldsymbol{\xi}} \\
     & =  - \int{q(\boldsymbol{\xi})\log p(\boldsymbol{D}|\boldsymbol{m}) d\boldsymbol{\xi}} 
    + \int{q(\boldsymbol{\xi})[\log q(\boldsymbol{\xi})- \log p(\boldsymbol{m})]d\boldsymbol{\xi}} +\log p(\boldsymbol{D})
      \label{eq:KL_rewritten}
\end{aligned}
\end{equation}

In this study, 
$q(\boldsymbol{\xi})$ and and $p(\boldsymbol{m})$ are assumed to be in normal distribution for simplicity. Namely,
\begin{equation}
\begin{aligned}
q(\boldsymbol{\xi}) &= \mathcal{N}(\boldsymbol{u}_{\boldsymbol{\xi}},\boldsymbol{\sigma}_{\boldsymbol{\xi}}) \\
p(\boldsymbol{m})  &= \mathcal{N}(\boldsymbol{u}_{\boldsymbol{m}},\boldsymbol{\sigma}_{\boldsymbol{m}})
\end{aligned}
  \label{eq:q_pm_definition}
\end{equation}
where $\boldsymbol{u}_{\boldsymbol{\xi}}$ and $\boldsymbol{\sigma}_{\boldsymbol{\xi}}$ are the desired vasculature and flow dynamics estimation and the uncertainty of estimation with given observations; $\boldsymbol{u}_{\boldsymbol{m}}$ and $\boldsymbol{\sigma}_{\boldsymbol{m}}$ are the mean and standard deviations of vasculature and flow dynamics parameters without given observations. Then, we can obtain
\begin{equation}
\begin{aligned}
&\int{q(\boldsymbol{\xi})\log q(\boldsymbol{\xi})d\boldsymbol{\xi}} = -0.5J\log(2\pi)- 0.5\sum_{j=1}^J{\log(\boldsymbol{\sigma}_{\xi(j)}^2+1)} \\
&\int{q(\boldsymbol{\xi})\log p(\boldsymbol{m})d\boldsymbol{\xi}} = -0.5J\log(2\pi)- 0.5\sum_{j=1}^J{\log(\boldsymbol{\sigma}_{\boldsymbol{m}(j)}^2+\frac{1}{\boldsymbol{\sigma}_{\boldsymbol{m}(j)}^2}(\boldsymbol{\sigma}_{\xi(j)}^2+(\boldsymbol{u}_{\xi(j)}-\boldsymbol{u}_{m(j)})^2))}
\end{aligned}
  \label{eq:q_pm_integral}
\end{equation}
where $J$ is number of elements/parameters in $\mu$ and $\sigma$, which is 4 for 3D ULM imaging. 

Following the reparameterization trick in conventional variational interference implementation, a posterior sample $\boldsymbol{w}$ for the estimated parameters is: $\boldsymbol{w}^{(l)}=\boldsymbol{u}_{\boldsymbol{\xi}}+\boldsymbol{\sigma}_{\boldsymbol{\xi}}\odot\boldsymbol{\epsilon}^{(l)}$, where $\boldsymbol{\epsilon}^{(l)}\sim p(\boldsymbol{\epsilon})=\mathcal{N}(\boldsymbol{0},\boldsymbol{I})$, $(l)$ denotes the $l^{th}$ time of sampling, and $\odot$ denotes element-wise production. Then, we assume the likelihood $p(\boldsymbol{D}|\boldsymbol{m})$ is also a Normal distribution, allowing us to define the optimization problem as minimizing the Mean Squared Error (MSE) between observed  MB speed and the MB speed predicted by the MB locations and the flow dynamic model in Eq. (1),
\begin{equation}
\begin{aligned}
p(\boldsymbol{D}|\boldsymbol{m}) &=  \frac{1}{\sqrt{2\pi}\sigma_e}exp(- \frac{\frac{1}{N} \sum_{n=1}^N(v^{(n)}-v_{\boldsymbol{w}^{(l)}}^{(n)})^2}{2\sigma_{e}^2})\\
v_{\boldsymbol{w}^{(l)}}^{(n)} &= G^\sharp_{\boldsymbol{w}^{(l)}} (R_{\boldsymbol{w}^{(l)}}^2-||\boldsymbol{x}^{(n)}-\boldsymbol{x}^c_{\boldsymbol{w}^{(l)}} ||^2) \\
\end{aligned}
  \label{eq:pDm_definition}
\end{equation}
where $N$ is number of observed data in $\boldsymbol{D}$ and $(n)$ denotes the $n^{th}$ trajectory; $\sigma_e$ is an assumed variance of the error between the observation and prediction. Then, the term in Eq. (\ref{eq:KL_rewritten}) can be estimated by Monte Carlo (MC) method as below,
\begin{equation}
\begin{aligned}
- \int{q(\boldsymbol{\xi})\log p(\boldsymbol{D}|\boldsymbol{m}) d\boldsymbol{\xi}} 
&= \log{(\sqrt{2\pi}\sigma_e)} +\int{p(\boldsymbol{\epsilon})( \frac{\frac{1}{N} \sum_{n=1}^N (v^{(n)}-v_{\boldsymbol{w}^{(l)}}^{(n)})^2}{2\sigma_{e}^2}) d\boldsymbol{\epsilon}} \\
& \overset{MC}{\approx} \log{(\sqrt{2\pi}\sigma_e)} +  \frac{1}{L} \sum_{l=1}^L( \frac{1}{2\sigma_{e}^2} \frac{1}{N} \sum_{n=1}^N (v^{(n)}-v_{\boldsymbol{w}^{(l)}}^{(n)})^2)
\end{aligned}
  \label{eq:pDm_integral}
\end{equation}
where $L$ is the number of MC sampling times. Introducing Eq.(\ref{eq:q_pm_integral}) and Eq.(\ref{eq:pDm_integral}) into Eq.(\ref{eq:KL_rewritten}), the equation can be rewritten as below,
\begin{equation}
\begin{aligned}
 \sigma_{e}^2\mathbf{KL}[q(\boldsymbol{\xi})||p(\boldsymbol{\xi})] &= \frac{1}{L} \sum_{l=1}^L( \frac{1}{2N} \sum_{n=1}^N (v^{(n)}-v_{\boldsymbol{w}^{(l)}}^{(n)})^2) \\
& + 0.5\sigma_{e}^2 \sum_{n=1}^J[-\log(\boldsymbol{\sigma}_{\xi(n)}^2+1) +
 \log(\boldsymbol{\sigma}_{\boldsymbol{m}(n)}^2+\frac{1}{\boldsymbol{\sigma}_{\boldsymbol{m}(n)}^2}(\boldsymbol{\sigma}_{\xi(n)}^2+(\boldsymbol{u}_{\xi(n)}-\boldsymbol{u}_{m(n)})^2))]
 \\&+ \sigma_{e}^2(\log p(\boldsymbol{D})+\log{(\sqrt{2\pi}\sigma_e)} )
\end{aligned}
  \label{eq:KL_rewritten_2nd}
\end{equation}

$\log p(\boldsymbol{D})$, $\log{(\sqrt{2\pi}\sigma_e)}$ and $\sigma_{e}^2$ can be regarded as constants in the minimization of the KL divergence \cite{bates2022probabilistic}. Therefore,

\begin{equation}
\begin{aligned}
q(\boldsymbol{\xi}) 
 =  \arg \min_{q(\boldsymbol{\xi})}  \quad \frac{1}{L} & \sum_{l=1}^L( \frac{1}{2N} \sum_{n=1}^N (v^{(n)}-v_{\boldsymbol{w}^{(l)}}^{(n)})^2) \\ 
 &+ 0.5\sigma_{e}^2 \sum_{n=1}^J[-\log(\boldsymbol{\sigma}_{\xi(n)}^2+1) 
+\log(\boldsymbol{\sigma}_{\boldsymbol{m}(n)}^2+\frac{1}{\boldsymbol{\sigma}_{\boldsymbol{m}(n)}^2}(\boldsymbol{\sigma}_{\xi(n)}^2+(\boldsymbol{u}_{\xi(n)}-\boldsymbol{u}_{m(n)})^2))]
\end{aligned}
  \label{eq:KL_rewritten_3nd}
\end{equation}
When getting the gradient of Eq. (\ref{eq:KL_rewritten_3nd}) respect to $\boldsymbol{u}_{\boldsymbol{\xi}}$ or $\boldsymbol{\sigma}_{\boldsymbol{\xi}}$ with the chain rule, the first line is related to the evidence and contributes to the gradient, which is equivalent to least square fit; the second line is not related to the evidence that can work as a regularization with a weight of $ \lambda=0.5\sigma_{e}^2$ and contribute to the gradient. 

The Adam optimizer is adopted to find the optimal $\boldsymbol{u}_{\boldsymbol{\xi}}$ and $\boldsymbol{\sigma}_{\boldsymbol{\xi}}$. The observation data $\boldsymbol{D}$ are normalized with zero-mean and unit variance. $\boldsymbol{u}_{\boldsymbol{\xi}}$ and $\boldsymbol{u}_{\boldsymbol{m}}$ are initialized and given as parameters obtained by the direct estimation with the normalized data, respectively. Then, Eq. (1) is rewritten in the format: $v=a\boldsymbol{x}^2+b\boldsymbol{x}+c$, as \{$a$, $b$, $c$\} are easier to converge than \{$G^\sharp=a$, $R$, $\boldsymbol{x}^c$\} in the Adam optimizer. Thus, the elements in $\boldsymbol{\sigma}_{\boldsymbol{\xi}}$ and $\boldsymbol{\sigma}_{\boldsymbol{m}}$ are corresponding to \{$a$, $b$, $c$\} in the SVI implementation.
$\boldsymbol{\sigma}_{\boldsymbol{\xi}}$ is initialized as a unit array.  The element in $\boldsymbol{\sigma}_{\boldsymbol{m}}$ corresponding to $a$, also the $G^\sharp$, is given as 10, and other elements are given as 100, which we empirically found helpful to reduce aggressive outliers under noisy condition and get acceptable estimation accuracy with $ \lambda=0.5\sigma_{e}^2 = 0.1$ from simulations.  The \emph{softplus} is used to force elements in $\boldsymbol{\sigma}_{\boldsymbol{\xi}}$ positive during the optimization. 
If the radius is estimated to be more than a limit that is empirically given as double or half the initial guess, the estimation is regarded as invalid to reduce outliers. When there is randomness in the MC sampling, the SVI estimation is repeated for 5 times with the each observation $\boldsymbol{D}$ and outputs the $\boldsymbol{u}_{\boldsymbol{\xi}}$ with the minimal \emph{det}(\emph{diag}($\boldsymbol{\sigma}_{\boldsymbol{\xi}}$)). 

\section{Method 2: Trajectory Grouping}
MB trajectories are grouped from ULM maps for SVI based on Hessian Analysis, as shown in  Algorithm \ref{alg:1}. Lines 1 to 5 are same with conventional Hessian analysis \cite{frangi1998multiscale}. Considering the sparsity of super-resolution map,  Algorithm \ref{alg:1} is implemented on non-zero voxels, denoted as $p$, that are a small portion of the map to save computation cost. When each voxel of the ULM speed map can be reconstructed by averaging the trajectory speed passing through the voxel, the number of trajectories in the average for the voxel, besides the speed value and position of the voxel, is also fed to the SVI and weights the average in Eq. \ref{eq:pDm_definition}, which approximates feeding individual trajectories to the SVI. Note that the half width of the center negative lobe of the second-order Gaussian derivatives $f_{G2}=\{f_{hh},f_{ww},f_{tt},f_{hw},f_{ht},f_{wt} \}$ is equal to its standard deviation, where subscripts $h,w,t$ is used to denote partial derivatives along rows, columns and slices of the 3D ULM speed map.

\begin{algorithm}
    \KwInput{ULM speed map $\boldsymbol{M}$, Searching range radius [$r_{min}$,$r_{max}$], radius size increment $r_{step}$}
    \KwOutput{Sample set $\boldsymbol{D}_{r,p}$ for each non-zero voxel in $\boldsymbol{M}_{v}$ and for each searching radius} 
    \For{ searching radius $r=r_{min}:r_{step}:r_{max}$}{
      $\cdot$  Give second-order Gaussian derivative function sets $f_{G2}$ with $\sigma=r$\\
       $\cdot$    Calculate Hessian matrix $\boldsymbol{H}$ for each pixel by the convolution between $\boldsymbol{M}_{v}$ and ${f_{G2}}$\\     
       $\cdot$   Get the eigenvalues and eigenvectors of $\boldsymbol{H}$ for each pixel\\
        $\cdot$  Rank the eigenvalues with its absolute from small to large and get $\boldsymbol{\lambda}_{H}$ for each pixel\\
        \For{Each non-zero pixel $p$ in $\boldsymbol{M_v}$}{
          \If{$\boldsymbol{\lambda}_{H}^{(2)}<0$ $\&$ $\boldsymbol{\lambda}_{H}^{(3)}<0$}{
            \For{cross-section center shift distance ratio $d_r= 0:0.25:1$}{
          $\cdot$  Find non-zero value set ${\{(\boldsymbol{x}, v)\}}_{r,p,d_r}$ from $\boldsymbol{M}_{v}$ within the cross section whose radius is $r$. \\
          \If{Meeting Justifications (1) $\&$ (2)}{ 
          $\cdot$ Execute SVI with the non-zero value set ${\{(\boldsymbol{x}, v)\}}_{r,p,d_r}$ \\
          \If{Meeting Justification (3)}{
            \textbf{return} Estimated Distributions of parameter \\ 
            \textbf{break}
            }
          }
          }
        }
    }
    }
    \caption{Grouping trajectories with vessel cross-section detection}
    \label{alg:1}
\end{algorithm}

For a voxel close to vessel edges, it is possible that the assumed $r$ approximates or is more than vessel radius but the cross section centered on the voxel can only capture very few samples; it is also possible that the cross section is across two close vessels and capture samples in two different vessels.
Therefore, in the implementation for line 10 and 12 in Algorithm \ref{alg:1},
we set three justifications: 
\\
(1) sample  numbers in $\boldsymbol{D}_{r,p}$ is $\geq N_{least} = 4$; $N_{least}$ is define as the unknown parameters in $\boldsymbol{m}$, i.e. 4 for 3D images. \\ 
(2) valid mean angle between trajectories within the cross section and norm of the cross section (A threshold is empirically given as 40 degree for the in vivo data).  \\
(3) valid radius (empirically defined within 0.5\emph{r} and 2\emph{r}) are estimated by SVI from $\boldsymbol{D}_{r,p}$. \\
When any of the above three is not satisfied, the section center is moved by a distance $d_{r}r$ and the ratio $d_{r}$ increases from 0 with a step of 0.25 until both justifications are met or $d_{r}=1$. The moving direction $\boldsymbol{v}_m$ is given as below
\begin{equation}
\begin{aligned}
&\boldsymbol{v}_m = \alpha_1 \boldsymbol{v}_{\lambda_{H}^{(2)}} + \alpha_2 \boldsymbol{v}_{\lambda_{H}^{(3)}} \\
&\alpha_1 = sign(\boldsymbol{v}_{J}\cdot\boldsymbol{v}_{\lambda_{H}^{(2)}})\sqrt{\lambda_{H}^{(2)2}/(\lambda_{H}^{(2)2}+\lambda_{H}^{(3)2})} \\
&\alpha_2 = sign(\boldsymbol{v}_{J}\cdot\boldsymbol{v}_{\lambda_{H}^{(3)}})\sqrt{\lambda_{H}^{(3)2}/(\lambda_{H}^{(2)2}+\lambda_{H}^{(3)2})}
\end{aligned}
  \label{eq:moving_direction}
\end{equation}
where $\lambda_{H}^{(2)}$ and $\lambda_{H}^{(3)}$ are the second and third elements of the eigenvalues; $\boldsymbol{v}_{\lambda_{H}^{(2)}}$ and $\boldsymbol{v}_{\lambda_{H}^{(3)}}$ are the corresponding normalized eigenvectors and coplanar with the cross section; the vector $\boldsymbol{v}_{J}$ is the first-order gradient of at $p$ obtained by convolving first-order Gaussian derivates $f_{G1}=\{f_{h}, f_{w},f_{t}\}$ with SR speed map $\boldsymbol{M}$; $\cdot$ denotes the inner product. With a suitable assumed radius $r$, the cross section can be moved close to the vessel center. 

\vspace{5mm}
\noindent\textbf{Computation Platform}

Codes for the proposed methods are developed in MATLAB R2024b. Data for simulation evaluation are processed on a desktop (CPU: Intel i9-14900k, Memory: 128 GB, GPU: NVIDIA 4070s-8G). Data for \emph{in vivo} demonstration are processed on another desktop (CPU: Intel U9-285k, Memory: 256 GB, GPU: NVIDIA RTX5060Ti-16G).

\newpage
\section{Supplementary Figures}
\begin{figure}[!hb]
    \centering
   \includegraphics[width=7 cm]{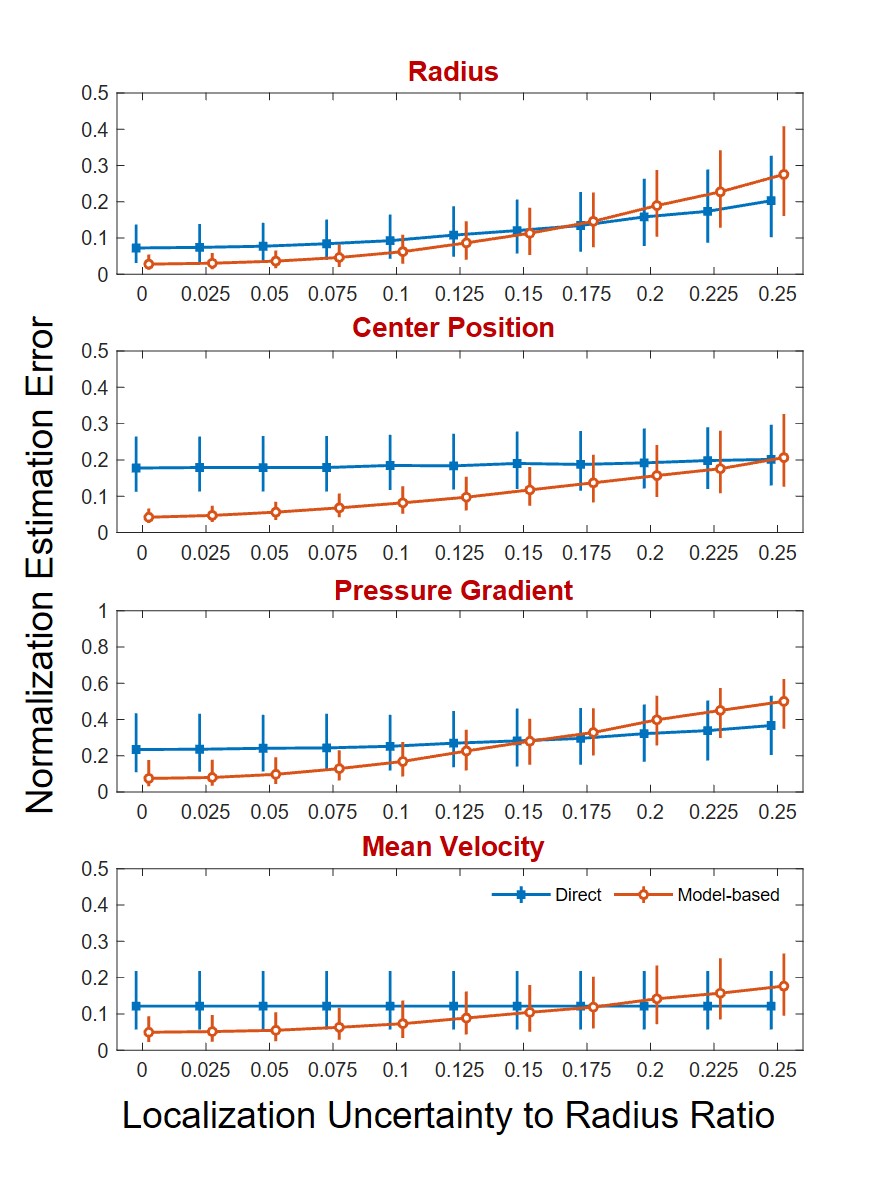}
    \caption{Impact of localization uncertainty on parameter estimation. Medians, 25th and 75th percentiles of estimation errors across all the simulated data samples for the four parameters obtained by the direct and model-based method versus different levels of localization uncertainty that is given by standard deviations of Gaussian noise. }
    \label{fig:S1}
\end{figure}

\newpage

\begin{figure}[!hb]
    \centering
   \includegraphics[width=10 cm]{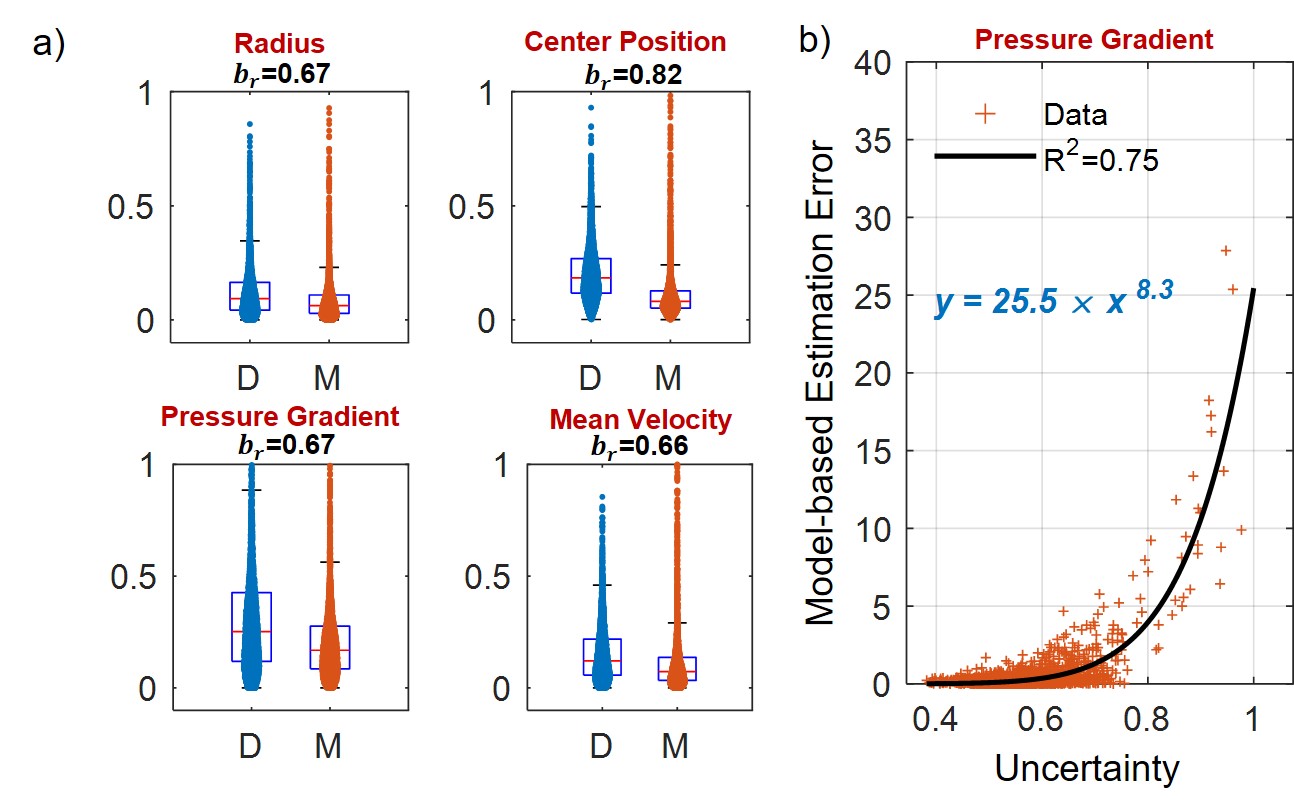}
    \caption{Parameter estimation with the ratio of localization uncertainty to radius as 0.1. a) direct estimation errors (denoted as 'D') versus model-based estimation errors (denoted as 'M'), where the proportion (denoted as '$b_r$') by which the latter is lower than the former is given on the top of the corresponding plot. There are 18 (different numbers of trajectories) $\times$ 200 (random sampling repetitions)= 3600 pairs, and Wilcoxon signed rank test implemented by MATLAB \emph{signrank} function returns $p$-value$<$0.001 for each of the four pairs. b) model-based estimation error versus estimation uncertainty with exponential fitting implemented.}
    \label{fig:S2}
\end{figure}
\newpage
\begin{figure}[!hb]
    \centering
   \includegraphics[width=11.5 cm]{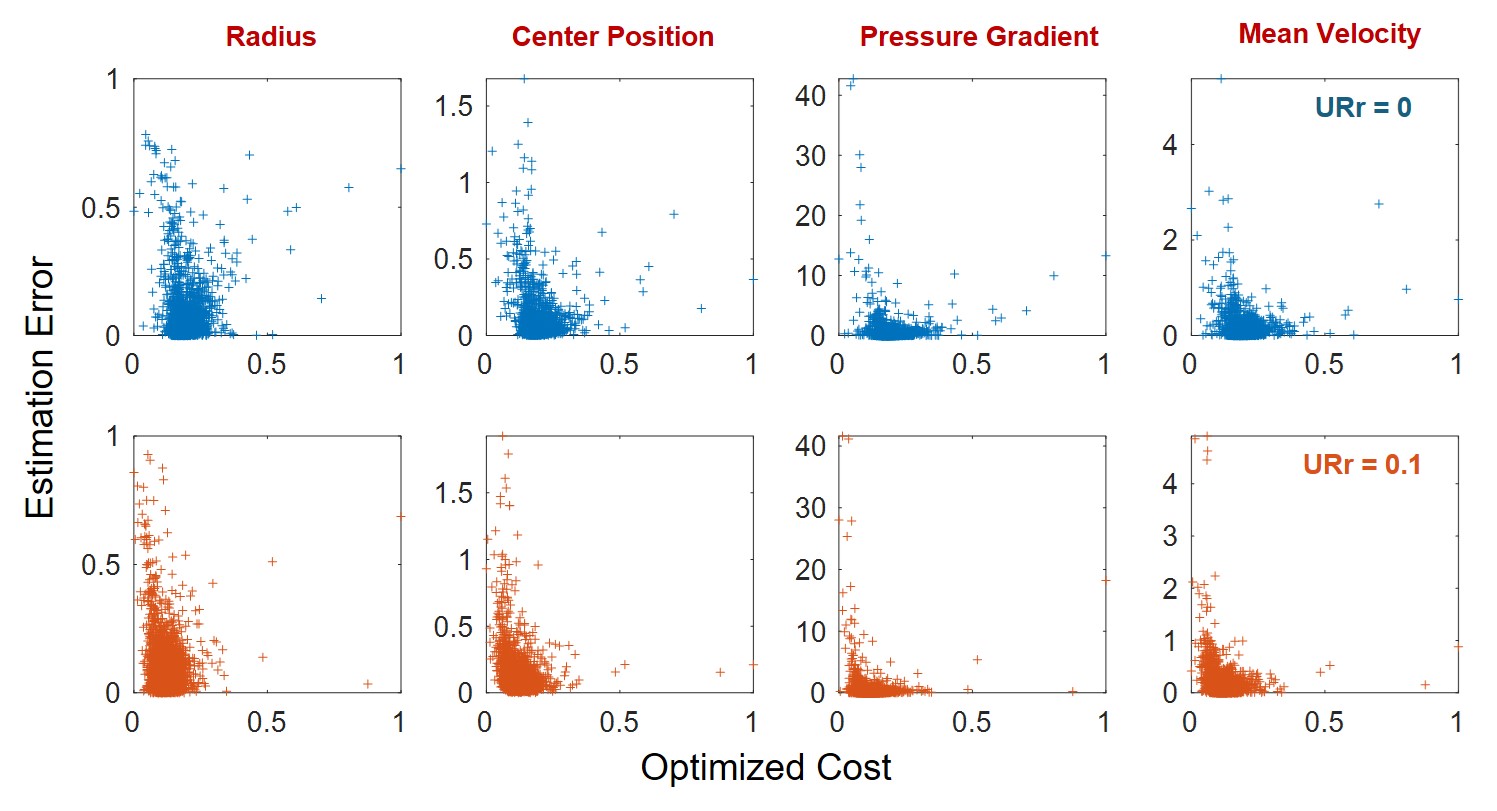}
    \caption{Optimized cost versus estimation error of the model-based method. URr denotes the ratio of localization uncertainty to radius ratio.}
    \label{fig:S3}
\end{figure}
\newpage
\begin{figure}[!hb]
    \centering
   \includegraphics[width=6.5 cm]{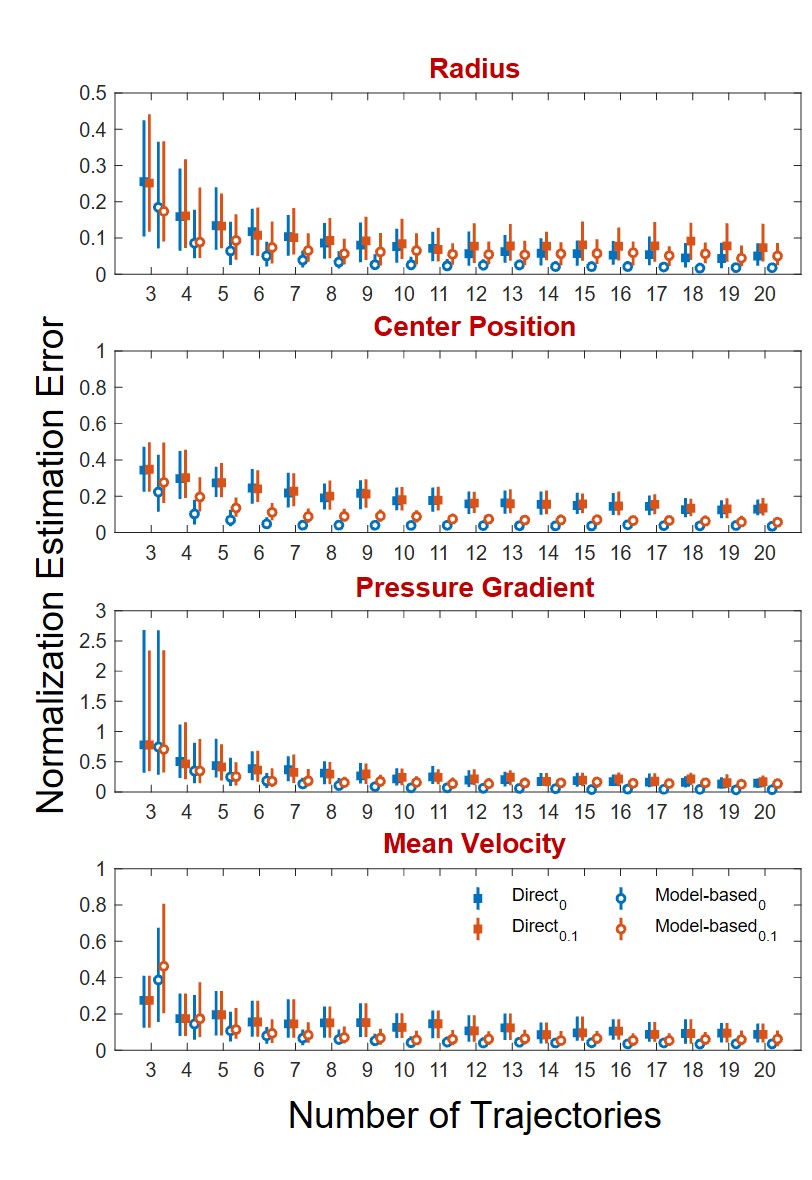}
    \caption{Impact of number of trajectories on parameter estimation. Medians, 25th and 75th percentiles of estimation errors obtained by the direct and model-based method versus number of trajectories in each sample group under two levels of localization uncertainty, where the subscript is the ratio of localization uncertainty to radius.}
    \label{fig:S4}
\end{figure}
\newpage

 \printbibliography